\documentclass[lettersize, journal]{IEEEtran}

\usepackage[caption=false,font=normalsize,labelfont=sf,textfont=sf]{subfig}
\usepackage{cite}
\usepackage{amsmath,amssymb,amsfonts}
\usepackage{algorithmic}
\usepackage{graphicx}
\usepackage{textcomp}
\usepackage{tabularx}
\usepackage{color, soul}
\usepackage{xcolor}
\usepackage{hyperref}
\def\BibTeX{{\rm B\kern-.05em{\sc i\kern-.025em b}\kern-.08em
		T\kern-.1667em\lower.7ex\hbox{E}\kern-.125emX}}
	
	\DeclareMathOperator{\E}{\mathbb{E}}
\begin{document}
	
\title{Online Self-Supervised {Deep} Learning for Intrusion Detection Systems \thanks{This research has been supported in part by the European Commission H2020 Program through the IoTAC Research and Innovation Action under Grant Agreement No. 952684 and by the European Commission Horizon Europe – the Framework Programme for Research and Innovation (2021-2027) DOSS Project under Grant Agreement No: 101120270.}
}

\author{Mert~Nak\i p and Erol~Gelenbe \IEEEmembership{Fellow, IEEE}
	\thanks{M. Nak\i p and E. Gelenbe are with Institute of Theoretical and Applied Informatics, Polish Academy of Sciences (PAN), Gliwice, Poland (e-mails: mnakip@iitis.pl, and seg@iitis.pl)}
	\thanks{E. Gelenbe is also with Lab. I3S, Universit\'{e} C\^{o}te d'Azur, Nice, France, and Ya\c{s}ar University, Izmir, Turkey}
	\thanks{This preprint is accepted for publication at IEEE Transactions on Information Forensics and Security, DOI: 10.1109/TIFS.2024.3402148. © 2024 IEEE.  Personal use of this material is permitted. Permission from IEEE must be obtained for all other uses, in any current or future media, including reprinting/republishing this material for advertising or promotional purposes, creating new collective works, for resale or redistribution to servers or lists, or reuse of any copyrighted component of this work in other works.}
}

\maketitle

\begin{abstract}
This paper proposes a novel Self-Supervised Intrusion Detection (SSID) framework, which enables a fully online {Deep Learning (DL)} based Intrusion Detection System (IDS) that requires no human intervention or prior off-line learning. The proposed framework analyzes and labels incoming traffic packets based only on the decisions of the IDS itself using an Auto-Associative Deep Random Neural Network, and on an online estimate of its statistically measured trustworthiness. The SSID framework enables IDS to adapt rapidly to time-varying characteristics of the network traffic, and eliminates the need for offline data collection. This approach avoids human errors in data labeling, and human labor and computational costs of model training and data collection. The approach is experimentally evaluated on public datasets and compared with well-known {machine learning and deep learning} models, showing that this SSID framework is very useful and advantageous as an accurate and online learning DL-based IDS for IoT systems.
\end{abstract}


\section{Introduction}
	
Botnet attacks can lead to thousands of infected devices \cite{Douligeris} compromising the devices of victims and turning them into ``bots'' via malware \cite{Goodin}, which in turn cause Distributed Denial-of-Service (DDoS) attacks. The \emph{malicious} bots, i.e. compromised devices, can generate fraud information, cause data leaks, and spread malware. It is reported that $27.7\%$ of all global website traffic in 2021 was generated by bots with malicious intent, and is growing with a $7.3\%$ increase reported between 2018 and 2021 \cite{bad_bots_2022}.

Botnet attacks severely challenge resource-constrained devices and Internet of Things (IoT) networks \cite{Tushir_impactsOfMirai}, as an attack propagates over the victim network increasing network congestion, power consumption, and processor and memory usage of IoT devices over time. Therefore, it is crucial to detect malicious network traffic and identify compromised IoT devices during an ongoing Botnet attack. While detecting malicious traffic allows reactive actions to alleviate the effects of the attack and stop it, identifying compromised IoT devices paves the way for preventive actions against the spread of malware and Botnet attack.

On the other hand, as the majority (approximately $52\%$) of IoT connections are to low cost and low maintenance devices deployed in massive IoT networks \cite{Cisco2020}, developing and implementing complex and advanced security methods is challenging as well. To this end, early research \cite{liu_survey} developed various types of lightweight Machine Learning (ML)-based Intrusion Detection Systems (IDS) --especially anomaly detecting IDS (anomaly-based IDS)-- for IoT networks, showing that anomaly-based IDS is very promising in detecting zero-day attacks based on unknown intrusions that often target vulnerable devices and networks. 

Since the decisions of anomaly-based IDS are highly dependent on the characteristics of the normal traffic used for parameter optimization (i.e. learning), accurate decisions become more difficult when the normal behaviour of network traffic changes over time due to both internal and external influences. For example, new device(s) may be added to the IoT network causing a considerable change in aggregated normal network traffic and an increased false positive alarms. 

Therefore, anomaly-based IDS could greatly benefit from the ability to adapt in real time to time-varying characteristics of network traffic, ideally through sequential online learning \cite{Incremental,CDIS}. However, the effectiveness of completely online learning, even for lightweight ML-based IDS, is often limited by two main factors: 1) A sufficient amount of collected and labeled traffic data is not always accessible for every system intended to be secured by the IDS. 
2) Online parameter updates are occasionally performed in parallel with intrusion detection at fixed or variable time intervals. Frequent online updates, i.e. short time intervals, result in high computational resource consumption for only minor or no performance gain per update. In contrast, infrequent updates, i.e. long time intervals, may have difficulty adapting to changes in normal traffic, resulting in poor IDS performance. Hence, each time interval needs to be carefully selected, taking into account the current state of the IDS and the actual behaviour of normal network traffic.

In this paper, in order to enable completely online learning of IDS parameters, a novel fully online Self-Supervised Intrusion Detection (SSID) framework is proposed. SSID learns from arriving traffic packets, measures the trustworthiness of the IDS, including its generalization ability and accuracy on traffic packets it uses for learning. It can then decide when to update the neural weights via its learning algorithm, keeping itself up-to-date with a high intrusion detection accuracy. 

The SSID framework can be used with any anomaly-based IDS that requires parameter optimization, providing fully online self-supervised learning of parameters in parallel with real-time detection requiring no human intervention. It also eliminates the need for labeled or unlabeled offline data collection, and offline training or parameter optimization. Therefore, the proposed framework contrasts sharply with much of existing work \cite{Song2021, Wang2021_byol, Zhang2022_self, Kye2022, Caville2022, Wang2023_self, Abououf2022, meyer2023federated} that has implemented self-supervised learning for intrusion detection, often utilizing offline (small-sized) labeled or unlabeled training data and pseudo-labeling. Accordingly, as its main advantages, the SSID framework
\begin{itemize}
	\item Enables IDS to easily adapt time varying characteristics of the network traffic, 
	\item Eliminates the need for offline data collection, 
	\item Prevents human errors in data labeling (online or offline), and 
	\item Avoids human labor and computational costs for model training and data collection through prior experiments.
\end{itemize}  

We also implement our SSID framework for a Deep Random Neural Network (DRNN)-based IDS that analyzes high-level network traffic metrics extracted from packet header information and learns those metrics calculated for only normal benign traffic. We evaluate the performance of the SSID framework for two tasks, malicious traffic detection and compromised device identification on Kitsune \cite{kitsune_dataset} and Bot-IoT \cite{botiot_dataset} datasets. The results revealed that IDS trained under the SSID framework achieve considerably high performance compared to the same IDS with offline and quasi-online (incremental and sequential) learning. Meanwhile, the IDS trained under SSID requires no offline dataset, external parameter optimization or human intervention.

The remainder of this paper is organized as follows: Section~\ref{sec:RelatedWorks} reviews the related work on intrusion detection. Section~\ref{sec:IDS} presents the overview of the IDS used in this work as well as the detection and learning processes. Section~\ref{sec:SSID} proposes the novel SSID framework and present the methodology enabling the self-supervised learning for IDS. Section~\ref{sec:Results} evaluates the SSID framework for malicious traffic detection and compromised device identification on public datasets, and compares its performance against the state-of-the art methods. Finally, Section~\ref{sec:Conclusion} summarizes this paper and provides some insights for the future work. {Note that the definitions of abbreviations and the symbols appear in this paper are respectively listed in Table~}\ref{table:abbreviation_SSID}{ and Table~}\ref{table:symbols_SSID}{ in Appendix.}


\section{Related Work}\label{sec:RelatedWorks}

We now briefly review recent related work on intrusion detection in three categories of the work that: 1) detect malicious traffic during Botnet-based DDoS (in short Botnet) attacks, 2) identifies compromised network nodes, and 3) performs self-supervised learning for intrusion detection.

\begin{table*}[t!]
	\centering
	\caption{{Key Features of Related Works on Self-Supervised Learning for Intrusion Detection}}
	\label{tab:comparison}
	\begin{tabular}{|c|c|c|c|c|}
		\hline
		\textbf{Reference} & \textbf{IDS} & \begin{tabular}[c]{@{}c@{}}\textbf{Labeled}\\ \textbf{Data}\end{tabular} & \begin{tabular}[c]{@{}c@{}}\textbf{Data} \\ \textbf{Generator}\end{tabular} & \begin{tabular}[c]{@{}c@{}}\textbf{Learning} \\ \textbf{Approach}\end{tabular} \\\hline
		
		Song and Kim \cite{Song2021} & \begin{tabular}[c]{@{}c@{}}Reduced Inception \\ ResNet\end{tabular} & \begin{tabular}[c]{@{}c@{}}Normal Traffic \\ +\\  Generated Noised Pseudo \\ Normal Data\end{tabular} & LSTM & Offline \\\hline
		
		Wang et al. \cite{Wang2021_byol} & \begin{tabular}[c]{@{}c@{}} BYOL Encoder \\+\\ Linear Classifier\end{tabular} & \begin{tabular}[c]{@{}c@{}} Malicious and Normal \\ Traffic\end{tabular} & BYOL & \begin{tabular}[c]{@{}c@{}} Offline \\ (Transfer Learning)\end{tabular} \\\hline
		
		Zhang et al. \cite{Zhang2022_self} & \begin{tabular}[c]{@{}c@{}} Deep Adversarial Anomaly\\  Detection (DAAD)\end{tabular} & \begin{tabular}[c]{@{}c@{}}Normal Traffic \\ +\\ Generated Feature Latents\end{tabular} & GAN & Offline \\\hline
		
		Kye et al. \cite{Kye2022} & \begin{tabular}[c]{@{}c@{}} AE-based Hierarchical \\ Anomaly Detection \end{tabular} & Only Normal Traffic & No Generator & Offline \\\hline
		
		Caville et al. \cite{Caville2022} & GNN & None & GraphSAGE & Offline \\\hline
		
		Wang et al. \cite{Wang2023_self} & AE-based Intrusion Score & None & Contextual Masking & Offline \\\hline
		
		Abououf et al. \cite{Abououf2022} & LSTM-AE & Only Normal Traffic & \begin{tabular}[c]{@{}c@{}}Auto \\ Encoder-Decoder\end{tabular} & Offline \\ \hline
		
		Meyer et al. \cite{meyer2023federated} & AE Neural Network&  \begin{tabular}[c]{@{}c@{}} Reduced Data of \\ Malicious and Normal \\ Traffic\end{tabular} & No Generator  & \begin{tabular}[c]{@{}c@{}} Offline \\ (Federated Learning)\end{tabular} \\\hline
		
		SSID Framework & AADRNN & None & No Generator & Online \\\hline
	\end{tabular}
\end{table*}

\subsection{DDoS Botnet Attack Detection}

In \cite{Tuan}, Tuan et al. conducted a comparative study for performance evaluation of ML methods aiming to classify Botnet attack traffic. In this work, the authors evaluated the performances of Support Vector Machine (SVM), MLP, Decision Tree (DT), Naive Bayes (NB), and unsupervised ML methods (such as K-means clustering) on two datasets (including KDD'99) revealing that unsupervised ML methods achieve the best performance with $98 \%$ accuracy. In \cite{Shao_adaptive_botnet}, Shao et al. created an ensemble of Hoeffding Tree and Random Forest (RF) models with online learning using both normal and attack traffic. In \cite{Shafiq}, Shafiq et al. developed a feature selection technique as a preprocessing algorithm for an ML-based botnet attack detector. This algorithm ranks features according to their Pearson correlation coefficients and greedily maximizes the detector's performance with respect to area under Receiver Operating Characteristic (ROC) curve in the Bot-IoT dataset. In \cite{Doshi}, Doshi et al. developed an attack detection algorithm comprised of feature extraction from the network traffic and ML classifier. In the place of the ML classifier, the authors used each of K-Nearest Neighbour (KNN), SVM, DT, and MLP methods; then, they evaluated the performance of this algorithm on a dataset collected within the same work. Letteri et al. \cite{Letteri} developed an MLP based Mirai Botnet detector specialized for Software Defined Networks. The authors fed 5 metrics, including the used communication protocol, to MLP. 

In \cite{Banerjee}, Banerjee and Samantaray performed experimental work to deploy a network of honeypots that attracts botnet attacks and to detect those attacks via ML methods, such as DT, NB, Gradient Boosting, and RF. In reference \cite{McDermott}, McDermott et al. developed the Bidirectional LSTM-based method which is developed for packet-level botnet attack detection by performing text recognition on multiple features including source and destination IP addresses of a packet. In addition, Tzagkarakis et al. \cite{Tzagkarakis} developed a sparse representation framework with parameter tuning using only normal traffic for botnet attack detection. 

Meidan et al. \cite{N_BaIoT} developed an ML-based attack detection technique which is trained using only normal traffic and tested for Mirai and Bashlite botnet attacks on an IoT network with nine devices. The authors also published the data collected in this study under the name N-BaIoT dataset. In order to detect Botnet attack in N-BaIoT dataset, Htwe et al. \cite{Htwe} used Classification and Regression Trees with feature selection, and Sriram et al. \cite{Sriram} performed a comparative study using 7 different ML methods (including NB, KNN, and SVM). In Reference \cite{Soe}, Soe et al. developed a Botnet attack detection algorithm comprised of two sequential phases first to train utilized ML method and perform feature selection, then to perform attack detection. The authors used MLP and NB within this architecture, and they evaluated the performance on N-BaIoT dataset. In \cite{Parra}, Parra et al. developed a cloud based attack detection method using Convolutional Neural Network (CNN) for phishing and using Long-Short Term Memory (LSTM) for Botnet attacks. The authors evaluated the performance of this method also on the N-BaIoT dataset achieving $94.8\%$ accuracy. CNN was also used by Liu et al. \cite{Liu_botnet} with features that are processed by the triangle area maps based multivariate correlation analysis algorithm. {In recent work }\cite{bovenzi2023network}, {Bovenzi et al. employed DL models, specifically using Auto Encoders (AEs) and KitNET, for unsupervised early anomaly detection in IoT datasets, namely IoT-23 }\cite{iot23dataset}{ and Kitsune} \cite{kitsune_dataset}. 
{The results of }\cite{bovenzi2023network}{ demonstrated the potential for early anomaly detection in IoT network attacks by evaluating the detection effectiveness of varying numbers of packets, finding the first four packets to be the most effective.}

\subsection{Compromised Device Identification}

Some recent work \cite{Kumar_MiraiBots,Chatterjee_bot,Nguyen_federated, Abhishek_gateway,Taneja, Prokofiev} focused on detecting compromised IoT devices during Botnet attacks, while  Kumar et al. \cite{Kumar_MiraiBots} detected Mirai-like bots scanning the destination port numbers in packet headers using an optimization-based technique for subsets of all IoT packets. Chatterjee et al. \cite{Chatterjee_bot} identified malicious devices in IoT networks via evidence theory-based analysis. To this end, they analyzed traffic flows and selected the rarest features from a large number of communication features, including number of connections, transport layer protocol, and source/destination ports. In \cite{Nguyen_Botnet_IIoT}, in order to detect IoT botnet in an Industrial IoT network, Nguyen et al. developed a dynamic analysis technique utilizing various ML models, such as SVM, DT, and KNN, based on the features generated from the executable files. 
In Reference \cite{Hristov}, Hristov and Trifonov developed a compromised device identification algorithm using wavelet transformation and Haar filter on the metrics indicating the processor, memory and network interface card usage of an IoT device. In \cite{Prokofiev}, Prokofiev et al. used Logistic Regression to determine if the source device is a bot based on 10 metrics regarding the traffic packets. The performance of logistic regression is tested for a botnet that spreads through brute-force attacks. Nguyen et al. \cite{Nguyen_federated} detected compromised devices by an anomaly detection technique combining federated learning with language analysis for individual device types identified prior to detection. In order to evaluate the performance of this technique, the authors collected a dataset by installing 33 IoT devices, 5 of which were malicious, and showed that detection performance is around $94 \%$ for positive and $99 \%$ for negative samples. 

More differently, in Reference \cite{Abhishek_gateway}, Abhishek et al. detected not compromised devices but compromised gateways monitoring the downlink channels in an IoT network and performing binary hypothesis test. In \cite{Trajanovski_framework}, Trajanovski and Zhang developed a framework consisting of honeypots to identify the indicators of compromised devices and botnet attacks. Bah\c{s}i et al. in \cite{Bahsi_dimensionality} addressed the scalability issues for ML-based Bot detection algorithms by minimizing the number of inputs of ML model via feature selection. In \cite{Taneja}, for mobile IoT devices, Taneja proposed to detect compromised devices taking into account their location, such that if a location change or current location of an IoT device is classified as unusual behavior, the device is considered compromised.

\subsection{Self-Supervised Learning for Intrusion Detection}

{Song and Kim }\cite{Song2021}{ developed self-supervised learning algorithm for anomaly-based IDS in in-vehicle networks. In this algorithm, Reduced Inception-ResNet is used to make binary classification and detect unknown (zero-day) attacks. The training of this anomaly-based IDS offline uses both normal traffic and noised pseudo data generated using LSTM. Wang et al. }\cite{Wang2021_byol}{ applied and adapted Bootstrap Your Own Latent (BYOL) self-supervised learning approach, which has been proposed in }\cite{BYOL}{, for intrusion detection. The parameters of the BYOL algorithm are learned, then updated via transfer learning, using a dataset containing both normal and malicious traffic samples. For anomaly detection, Zhang et al. } \cite{Zhang2022_self}{ developed deep adversarial training architecture by extending the well-known bidirectional Generative Adversarial Network (GAN) model. This architecture jointly learns from normal data and generated latent features. Kye et al. }\cite{Kye2022}{ introduced a hierarchical network IDS based on the Auto Encoder (AE). By leveraging self-supervised signals and specialized anomaly scores within its AE architecture, this IDS learns offline only from normal traffic data and does not need an additional data generator. Caville et al. }\cite{Caville2022} {developed a Graph Neural Network (GNN) based network-level IDS. This IDS was trained with a self-supervised learning approach using both positive and negative samples for offline training with an encoder graph created using an extended version of the well-known GraphSAGE framework. Wang et al.} \cite{Wang2023_self}{ developed an IDS with unsupervised learning combined with transformer based self-supervised masked context reconstruction, which improves the learning by magnifying the abnormal intrusion behaviours. Abououf et al. }\cite{Abououf2022}{ developed a lightweight IDS architecture based on LSTM Auto Encoder (LSTM-AE) to perform detection on IoT nodes. This model is trained offline and unsupervised in an encoder-decoder architecture using a pre-collected dataset in the cloud. Meyer et al. }\cite{meyer2023federated}{ developed a federated self-supervised learning IDS. This IDS employ auto-encoder based self-supervised model on local datasets using federated approach. Note that the key features of the related works on self-supervised learning, as well as the proposed SSID framework, are summarized in Table~}\ref{tab:comparison}.


{In this paper, we introduce a pioneering learning framework termed SSID, tailored for Deep Learning (DL)-based intrusion detection. Unlike conventional approaches prevalent in the {ML / DL} literature, SSID distinguishes itself in several key aspects:}
\begin{itemize}
	\item {SSID eliminates dependency on additional generative models. Unlike common practice in self-supervised learning, which often necessitates the incorporation of an extra generative model (or contrastive method) for training purposes} \cite{Liu_SSL_survey}{, our SSID framework operates without this requirement.}
	
	\item {SSID is independent of offline training data. While many methods in the literature rely on the availability of pre-collected (unlabeled) offline training data }\cite{Yu_SSL_survey}{, SSID is designed to function autonomously, even in scenarios where such data is not readily accessible. However, if available, SSID can leverage offline data to enhance its performance.}
	
	\item {SSID facilitates fully online learning on independent network nodes. A distinctive feature of SSID is its ability to facilitate continuous online learning and enable real-time intrusion detection across independent network nodes.}
\end{itemize}

\section{Intrusion Detection System Used in SSID}\label{sec:IDS}

The SSID framework does not consider a specific algorithm for IDS or have strict requirements for it, except that it is based on ML / DL or some other function with learnable parameters and has a certain range of inputs and outputs. In addition, the SSID framework can also be used with both anomaly and signature based detection algorithms. On the other hand, as real-time network traffic contains only normal ``benign'' traffic until an attack occurs, an IDS structure that can learn only from normal traffic 
may provide higher performance under self-supervised learning. Therefore, anomaly-based algorithms are the main focus in this paper.  

\begin{figure}[h!]
	\centering
	\includegraphics[scale=0.8]{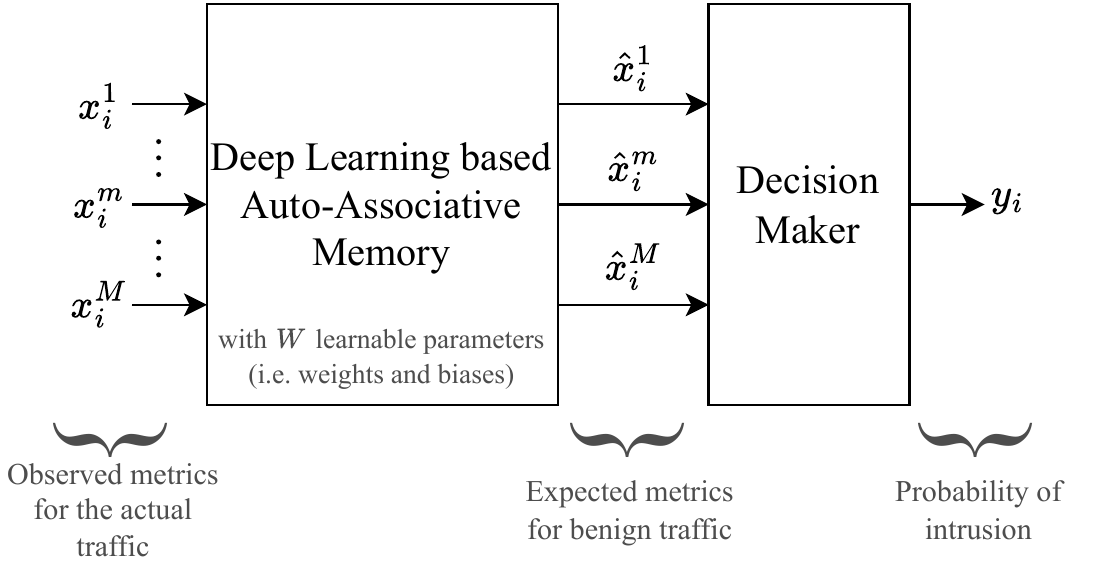}
	\caption{Particular structure of IDS used within the SSID framework during performance evaluation}
	\label{fig:ids_example}
\end{figure}

We first present the structure of anomaly-based IDS that we used within the SSID framework. This particular IDS structure is displayed in Figure~\ref{fig:ids_example}, which is mainly comprised of an DL model and a decision maker component. 



In the IDS structure shown in Figure~\ref{fig:ids_example}, the DL model is used to create an Auto-Associative Memory (AAM) that reconstructs benign traffic metrics --which are the expected metrics according to the norm of the actual traffic learned by the AAM-- from observed metrics. The significant difference between the reconstructed (expected) metrics and the actual metrics, as measured by the decision maker component, may be indicative of malicious traffic.

The input of this IDS is the vector of $M$ network traffic metrics, $x_i = [x_i^m, \dots, x_i^m, \dots, x_i^M]$, and the output, $y_i$ is a decision indicating the probability of intrusion corresponding to current traffic. The vector of expected metrics, which is the output of DL-based AAM, for packet $i$ is denoted by $\hat{x}_i = [\hat{x}_i^1, \dots, \hat{x}_i^m, \dots, \hat{x}_i^M]$. The DL-based AAM is a learned function that maps the noisy or disordered metrics to the normal metrics, i.e. $f_{aam}: x_i \mapsto f_{aam}(x_i)$ for $f_{aam}(x_i) = \hat{x}_i$, so that $f_{aam}: [0, 1]^M\to [0, 1]^M$. 

\subsection{Deep Random Neural Network as Auto-Associative Memory}

{In the particular implementation of the IDS shown in Figure~}\ref{fig:ids_example}{, in place of the DL algorithm, we use the DRNN model }\cite{Deep1,Deep2}{.} DRNN is an extension of the RNN \cite{RNN,RNN2} with dense feedback loops between clustered neuronal somas, and an overall feed-forward structure between layers of dense clusters. After it is trained to create an AAM, we call it Auto-Associative DRNN (AADRNN). Earlier research has shown that an AADRNN-based IDS has a lightweight architecture and offers high accuracy when used with unsupervised auto-associative training with normal (benign) traffic \cite{Brun,Brun2, MIRAI, Incremental,CDIS, G_network}. The AADRNN-based IDS was also evaluated with offline \cite{MIRAI}, incremental \cite{Incremental} and sequential \cite{CDIS} learning to detect malicious traffic and compromised devices during Botnet attacks. G-Networks \cite{GNETS}, which generalize the RNN, and the simple RNN itself were also used with offline learning to detect zero-day \cite{G_network} and SYN DoS \cite{Evmorfos} attacks. {In contrast, in this paper, we use the AADRNN-based IDS as a specific IDS within the novel SSID framework proposed for the first time in this paper. The parameters of the AADRNN-based IDS are now learned online thanks to the SSID framework learning approach that is completely online and does not require labeled data. It should be noted that the main contribution of this paper is the development of self-supervised approach and the novel SSID framework. In addition, AADRNN-based IDS can be replaced with another ML or DL based IDS, e.g. MLP will also be evaluated within the SSID framework in Section~}\ref{sec:Results}{.}


In the AADRNN-based IDS, we use a DRNN model consists of $H$ layers and a rectangular structure with equal units at each layer. Accordingly, we set $H=M$, each hidden layer $h \in \{1, \dots, H-1\}$ contains $M$ neural clusters, and the output layer is comprised of $M$ linear neurons.


\subsection{Intrusion Detection Process}

Recall that this paper considers two main tasks: detecting malicious traffic packets and identifying compromised network nodes. As the results of earlier research \cite{MIRAI, CDIS} showed, each task requires a customized IDS to provide high accuracy. Therefore, in the remainder of this subsection, we present the traffic metrics and the decision maker component for each task separately. 

\subsubsection{Traffic Metrics}
{We use the traffic metrics that are defined in our earlier research specifically for Botnet attack detection in }\cite{MIRAI} {and compromised device identification in }\cite{CDIS}{. These metrics are relatively few in number and are calculated based solely on packet header information. Therefore, they remain anonymous regarding packet content and communicating devices, do not need for any sensitive or device-specific information, thus preventing IDS from making biased decisions, and are suitable for real-time operation on lightweight systems. Meanwhile, they have been shown to be effective in capturing signatures of Botnet attacks.}
	

\textbf{For malicious traffic detection}, in order to capture the signatures of Botnet attacks (especially Mirai), as the inputs of AAM, we use $M = 3$ normalized metrics that measure the total size and average inter-transmission times of the last $500$ packets and the number of packets in the last $100$ seconds.

\textbf{For compromised device identification}, we use $M=6$ normalized traffic metrics which are calculated over time windows of length $10$ seconds. These metrics measure the average size and average number of packets received from a single source, the maximum size and maximum number of packets received from any single source, and the total size and total number of packets transmitted during a time window.

\subsubsection{Estimation of the Expected Benign Traffic Metrics}

The expected benign traffic metrics is estimated (i.e. reconstructed) based on the traffic metrics given as the inputs of AADRNN. The estimated traffic metrics ($\hat{x}_i$) are then fed into the decision maker component. 

Let $W_h$ denote the $[(M+1) \times M ]$ matrix of connection weights (including biases) between the layer $h-1$ and layer $h$ for $h \in \{1, \dots, H\}$; that is, $W_h$ is the multiplier for the inputs of layer $h$. In addition, $\zeta(\cdot)$ denote the activation function of a cluster in AADRNN. 

Accordingly, in real-time operation, the forward pass of AADRNN model for the given input vector $x_i$ is computed as:
\begin{align}
&\hat{x}_{(i, 1)} = \zeta([x_i, 1] \, W_{1})\label{eq:forward1_DRNN}\\
&\hat{x}_{(i, h)} = \zeta([\hat{x}_{(i, h-1)}, 1] \, W_{h}) \quad \forall h \in \{2, \dots, H-1\},\label{eq:forward_DRNN}\\
&\hat{x}_i = [\hat{x}_{(i, H-1)}, 1] \, W_{H},\label{eq:output_DRNN}
\end{align}
where $\hat{x}_{(i, h)}$ is the output of layer $h$ for packet $i$, and the term $[x_i, 1]$ or $[\hat{x}_{(i, h)}, 1]$ indicates that $1$ is added to the input of each layer as a multiplier of the bias. 

The activation function $\zeta(\cdot)$ is defined as \cite{Deep1}:  
\begin{eqnarray}
	\zeta(\Lambda) &=& \frac{p \, (r + \lambda^+) + \lambda^- + \Lambda }{ 2 \, [\lambda^- + \Lambda ]}\\
	& - &\sqrt{ {\Big( \frac{p \, (r + \lambda^+) + \lambda^- + \Lambda }{ 2 \, [\lambda^- + \Lambda ]} \Big)}^2 - \frac{\lambda^+}{\lambda^- + \Lambda }}~, \nonumber
\end{eqnarray}
where $\Lambda$ is the input of the given cluster, $p$ is the probability that any neuron received trigger transmits a trigger to some other neuron, and $\lambda^+$ and $\lambda^-$ are respectively the rates of external Poisson flows of excitatory and inhibitory input spikes to any neuron.

\subsubsection{Decision Making}

{The last step in the attack detection process is making the attack decision. Since the AADRNN output provides the expected metrics for normal network traffic, any deviation of the actual metrics from the AADRNN output, i.e. the expected metrics, indicates malicious traffic. Therefore, we use a lightweight decision making approach based on comparison of actual and expected metric values. In particular, we measure the deviation of the actual metrics $x_i$ from the expected metrics $\hat{x}_i$. On the other hand, since AADRNN-based IDS uses different set of metrics for malicious traffic detection and compromised device identification tasks, the decision making approach is also slightly different for each task. One should note that this lightweight decision making approach requires no human intervention or parameter settings based on offline data; therefore, it is suitable for online learning IDS.} 

\textbf{For malicious traffic detection}, {the decision maker calculates the output of IDS, $y_{i}$, as the average absolute difference between the actual and the expected metric values:}
\begin{equation}
	y_i = \frac{1}{M}\sum_{m=1}^{M}|x_i^m - \hat{x}_i^m|.
\end{equation}

\textbf{For compromised device identification}, {the decision maker calculates the output of IDS as the maximum absolute difference between the actual and the expected metric values:}
\begin{equation}
	y_{i} = \max_{m\in \{1, \dots, M \}}(|x_{i}^m - \hat{x}_{i}^m|).
\end{equation}

\subsection{Learning Algorithm to Create AADRNN}


In order to create an AADRNN, we train an DRNN model based on a batch of packet samples, denoted by $B^l$, collected and provided by the proposed SSID framework at any learning phase $l$. To this end, we use the learning algorithm, used for DRNN in earlier research \cite{MIRAI, Incremental, CDIS}, which shall be reviewed in this subsection.


\subsubsection{Initial Learning}

As we use sequential learning to create an auto-associative memory from DRNN (namely, AADRNN), each weight matrix $W_h$ is learned and updated based on only the batch of benign packets $B^l$ provided by the SSID framework.


When the learning algorithm is called by SSID for $B^l$ at initial learning phase $l=0$, we solve the following problem using A Fast Iterative Shrinkage-Thresholding Algorithm (FISTA) \cite{FISTA}: 
\begin{align}\label{eq:minimization_sequential}
	W_h = \arg\hspace{-0.5cm}\min_{\{W: \, W\geq \, 0\}} \, \Bigl( \, \Bigl|\Bigl| [ & adj(\zeta(\hat{X}^\text{train}_{l, h-1} W_{R~})), \mathbf{1}_{k \times 1}] \, W  \nonumber\\ 
	&- \hat{X}^\text{train}_{l, h-1} \, \Bigr| \Bigr|_{L_2}^2 + ||W||_{L_1} \, \Bigr)~,
\end{align}
where $W_R$ is a random weight matrix whose elements are in $(0, 1)$ and $\hat{X}^\text{train}_{l, h}$ is the matrix that collects the output vectors of layer $h$ for all packets $i \in B^l$ if $y_i = 0$:
\begin{equation}
	\hat{X}^\text{train}_{l, h} = \{\hat{x}_{(i, h)}, ~ \forall i \in B^l\}, 
\end{equation}
\begin{equation*}
	\hat{X}^\text{train}_{l, 0} = \{x_i, ~ \forall i \in B^l\},~~ \text{and} ~~ \hat{X}^\text{train}_{l, H} = \{\hat{x}_i, ~ \forall i \in B^l\}
\end{equation*}
After FISTA is performed for a predefined number of iterations to solve (\ref{eq:minimization_sequential}), we normalize the resulting weight matrix $W_h$: 
\begin{equation}
	W_h \leftarrow 0.1\frac{W_h}{\max\big(\hat{X}^\text{train}_{i, h}\big)}~.
\end{equation}

\subsubsection{Online Incremental Learning}

During the online incremental learning of IDS parameters (i.e. $l \geq 1$), only the connection weights of the output layer of AADRNN, $W_H^l$, are updated via the second stage to learn the new patterns of the benign traffic provided by SSID via $B^l$. As the incremental learning algorithm, we mainly integrate the sequential learning algorithm developed in \cite{sequential_learning}. Let define the operation matrix $O_l$, which is initialized for $l=0$ as the inverse of the Gram matrix:
\begin{equation}
	O_0 = \Bigl[(\hat{X}^\text{train}_{l, H})^T \, \hat{X}^\text{train}_{l, H}\Bigr]^{-1}
\end{equation}
If there is at least one benign packet in $B^l$, we first compute the value of $O_l$ for $l \geq 1$:
\begin{align}\label{eq:P_update_incremental}
	O_l = O_{l-1} - &  O_{l-1} (\hat{X}^\text{train}_{l, H-1})^T   \\\nonumber 
	& \Bigl[I + \hat{X}^\text{train}_{l, H-1} O_{l-1} (\hat{X}^\text{train}_{l, H-1})^T \Bigr]^{-1} \hat{X}^\text{train}_{l, H-1} O_{l-1}
\end{align}
Then, the value of $W_H$ is updated: 
\begin{equation}\label{eq:weights_update_incremental}
	W_H \leftarrow W_H + O_l (\hat{X}^\text{train}_{l, H-1})^T (\hat{X}^\text{train}_{l, 0} - \hat{X}^\text{train}_{l, H-1} W_H)
\end{equation}

\subsubsection{Learning Error Provided to the SSID Framework}

Furthermore, since we use an anomaly-based algorithm that learns only the benign traffic, we take the learning error as the mean of estimated attack probabilities for packets in learning batch $B^l$: 
\begin{equation}
	\mathcal{E}(l) = \frac{1}{|B^l|} \sum_{i \in B^l} {y_i}
\end{equation}





\section{The Self-Supervised Intrusion Detection Framework}\label{sec:SSID}

\begin{figure}[h!]
	\hspace{-0.3cm}
	\includegraphics[scale=1]{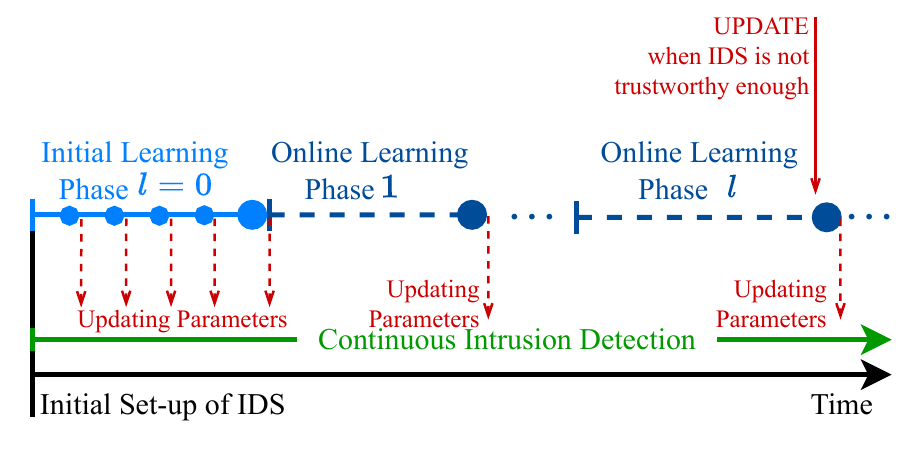}
	\caption{Detection and learning processes of IDS within the Fully Online Self-Supervised Intrusion Detection (SSID) framework}
	\label{fig:system_diagram}
\end{figure}

\begin{figure*}[t!]
	\centering
	\includegraphics[scale=0.6]{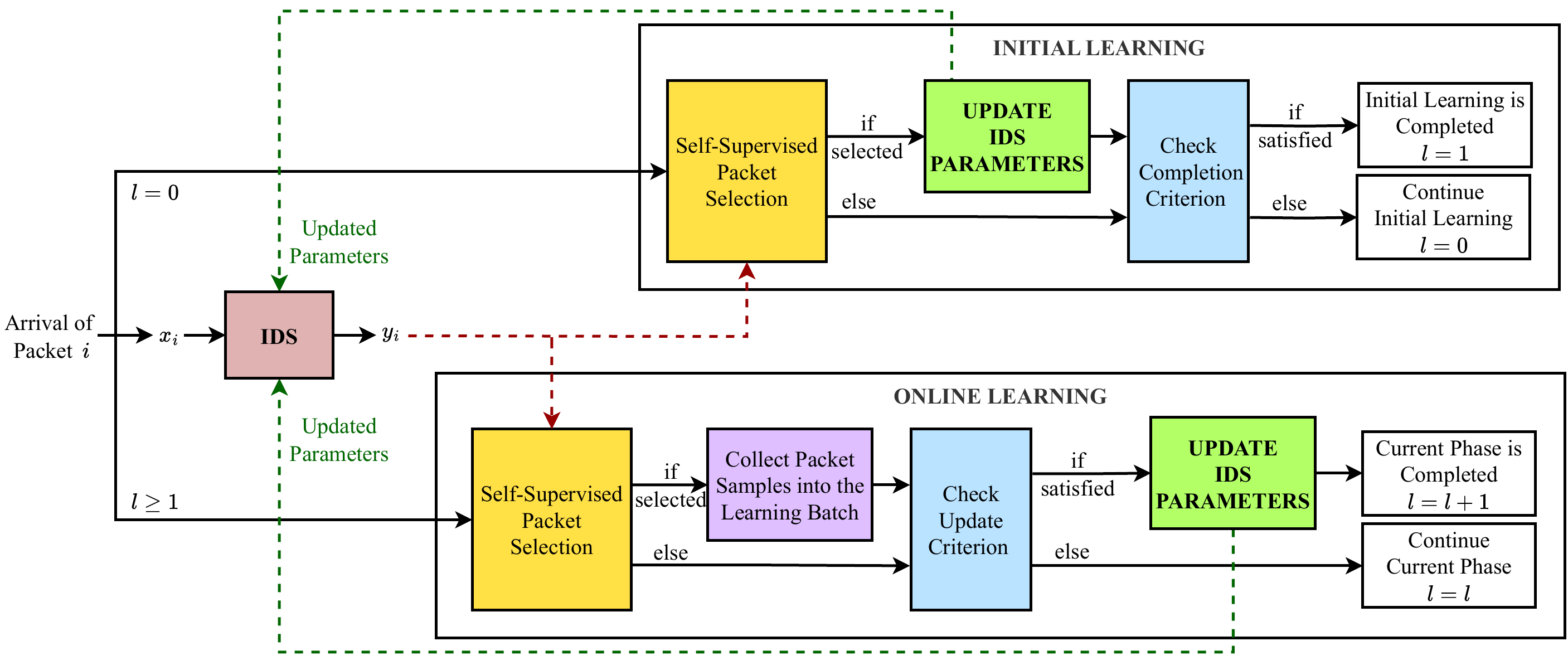}
	\caption{Block diagram of the learning process in the SSID framework for online self-supervised learning of the parameters of IDS}
	\label{fig:SSID_framework}
\end{figure*}

As the main contribution of this paper, we propose the novel SSID framework to enable fully online self-supervised learning of the parameters of IDS with no need for human intervention. 
In order to clearly present the SSID framework, this section first explains the main functionalities of both the initial and online learning phases through Figure~\ref{fig:system_diagram}. Then, the comprehensive methodology, that includes the details regarding all the blocks in Figure~\ref{fig:SSID_framework}, is presented. 

As shown in Figure~\ref{fig:system_diagram}, within the SSID framework, there are two main operations performed, {intrusion detection (lower line) and learning (upper line)}. Intrusion detection is the main operation performed by IDS and is not modified by SSID. That is, intrusion detection (as an operation) is defined only by a particular IDS algorithm used in SSID. 
On the other hand, we can say that our SSID framework ensures that IDS makes accurate decisions by updating its parameters with online self-supervised learning, and it performs intrusion detection uninterruptedly and continuously. {Regarding the communication (data transfer) between intrusion detection and online self-supervised learning processes in SSID, it is important to note that SSID uses the decisions of IDS to enable self-supervised learning during either initial or online learning phases.}




\subsection{Online Self-Supervised Learning}\label{sec:learning_process}

In parallel with attack detection, our SSID framework provides online self-supervised learning of IDS parameters, as shown on the upper process line in Figure~\ref{fig:system_diagram}. As seen in that figure and Figure~\ref{fig:SSID_framework} which shows the learning process in SSID, the online self-supervised learning process starts with the initial learning phase (namely, $l=0$) and continues with successive online learning phases.

{Since network traffic characteristics may vary with time and substantially affect the IDS detection performance, it is crucial to update the parameters of the IDS online concurrently with its real-time operation. The parameter updates conducted through online learning serve to enhance the reliability of the IDS as a detector by enhancing its performance and ensuring it remains aligned with the latest traffic characteristics. Online learning also circumvents the need to collect and label extensive datasets for offline training, thereby conserving both time and resources. In situations where labeled datasets are readily accessible, the IDS can be pre-trained, and its performance can be validated using those datasets.}

 

\subsubsection{Initial Learning}

The initial learning phase in SSID can be considered a special case of the proposed methodology of self-supervised learning, which allows IDS to be used from its initial setup and updates the parameters of the IDS frequently achieving the desired performance gradually and quickly. 

In detail, as shown in the top block of Figure~\ref{fig:SSID_framework}, during the initial learning phase in SSID, the parameters of the IDS are updated 
for each packet that is selected for learning via our self-supervised packet selection methodology. Whether the parameters of the IDS are updated or not, SSID checks the trust based completion criterion of the initial learning phase $l=0$ aiming to complete this phase as soon as the IDS is trained enough to make trustworthy decisions. To this end, it first calculates the trustworthiness of the IDS, namely the ``trust coefficient'' denoted by $\Gamma \in [0, 1]$, which indicates the confidence of SSID in any decision made by the IDS. Since the IDS does not have any information about the network traffic patterns yet, SSID cannot judge the decisions of the IDS and starts the initial learning process with $\Gamma = 0$ meaning that there is no trust in the decisions of the IDS.

Subsequently, SSID checks the trust criterion to complete the initial learning phase $l=0$ by measuring if SSID's trust in the IDS is greater than a threshold $\Theta$, is the minimum desired trust level:
\begin{equation}\label{eq:criterion_complete}
	\text{if} ~ \Gamma \geq \Theta, ~\text{complete initial learning and set } l=1
\end{equation}\\
Thus, {as given in }(\ref{eq:criterion_complete}), if $\Gamma \geq \Theta$, the initial learning phase has been completed, and the next packet will be considered for the first phase $l=1$ of continuous online learning.


\subsubsection{Online Learning}

After the initial learning is completed, the parameters of IDS are updated via an online learning phase $l \geq 1$ when the trust of SSID in the IDS is unacceptably low. 
As the lower block in Figure~\ref{fig:SSID_framework} shows, the parameters of the IDS are updated for a collected batch of packets when the trustworthiness of the IDS is not acceptable anymore. When SSID is in the online learning phase $l\geq 1$, each packet $i$ selected by our self-supervised packet selection method is collected into the batch of training packets, $B^l$. 

Then, SSID checks the trust-based criterion to update the parameters of the IDS. Inversely with the initial learning phase, SSID now updates the parameters of the IDS if $\Gamma < \Theta$, at least $K$ packets are collected for learning (i.e. $|B^l| \geq K$), and there is no attack detected by the IDS:
\begin{align}\label{eq:criterion_update}
	&\text{if} ~ \Gamma < \Theta ~\text{and}~ |B^l| \geq K ~\text{and}~ \frac{1}{I}\sum_{j=i-I+1}^{i} y_j \leq \gamma, \nonumber \\
	&\text{update parameters and set } l=l+1
\end{align}\\
where $I$ is the number of packets to calculate the average of the intrusion decisions, $\gamma$ is the intrusion threshold, and $K$ is provided by the user considering properties of the network and learning algorithm. Limit of minimum $K$ packets is added only to provide practical efficiency for training.

That is, {by (}\ref{eq:criterion_update}{), }SSID waits for a considerable decrease in the trustworthiness of the decisions of IDS to update the parameters since $\Gamma$ is known to be already greater than $\Theta$ at the end of the initial learning phase $l=0$. In this way, the learning is performed when it is essential. 

On the other hand, if an intrusion is detected where the average output of the IDS is greater than $\gamma$, SSID clears the batch of collected packet samples, $B^l$:
\begin{equation}\label{eq:clearance}
	\text{if} ~ \frac{1}{I}\sum_{j=i-I}^{i} y_j > \gamma, ~\text{empty} ~B^l
\end{equation}
With this cleanup, SSID aims to prevent the IDS from learning any false negative instances since false negative outputs are very likely to occur just before an attack is detected.

\subsection{Self-Supervised Packet Selection}\label{sec:selecting_packets}

In the remainder of this section, we present our proposed methodology to train the utilized IDS in a self-supervised fashion enabling the fully online property of SSID. In other words, we explain the details of the learning process in SSID, which are shown as subblocks in Figure~\ref{fig:SSID_framework}. 

{As the first operation of the learning process in SSID, each packet $i$ is decided to be used in learning for the next update of IDS parameters.} 
The packet selection is executed in a self-supervised manner that only considers the output of the IDS together with SSID's trust in it.

Let $p_i^-$ and $p_i^+$ respectively be the probability of selecting packet $i$ to be used as a \textit{benign} or \textit{malicious} packet sample in the training of IDS, and $q_i$ be the probability of rejecting $i$ to be used in training. That is, we select the packet $i$ as the sample of a benign packet with probability $p_i^-$ or that of an attack packet with probability $p_i^+$ to use it in training, or the packet $i$ is not included in the training set with probability $q_i$. Also, recall that $y_i \in [0, 1]$ is the output of IDS for packet $i$.

Since we assume that there are no packet labeling mechanisms or labor to prepare packet data for learning, we select each packet $i$ based on the output of IDS (which is the estimation of the probability of packet $i$ being malicious) considering how trustworthy IDS is. Therefore, we shall also define a trust coefficient $\Gamma$ to measure the trustworthiness of IDS at any time based on the representativeness of the packet samples that IDS learned until the end of the last learning phase and the generalization ability of IDS from these samples. 

Accordingly, we start by defining $p_i^+$ as 
\begin{equation}
	p_i^+ \equiv \textrm{(trust in IDS) (est. prob. of packet $i$ being malicious)} \nonumber
\end{equation}
\begin{equation}
	p_i^+ = \Gamma \, y_i
\end{equation}
We further define $p_i^-$ similarly to $p_i^+$:
\begin{equation}
	p_i^- \equiv \textrm{(trust in IDS) (est.  prob. of packet $i$ being normal)} \nonumber
\end{equation}
\begin{equation}
	p_i^- = \Gamma \, (1-y_i) 
\end{equation}
Subsequently, since
\begin{equation}
	p_i^+ + p_i^- + q_i = 1,
\end{equation}
the probability $q_i$ of not selecting the packet $i$ for training is:
\begin{eqnarray}
	q_i &=& 1 - (p_i^+ + p_i^-) \nonumber\\ 
	&=& 1 - \Gamma 
\end{eqnarray}

Recall that SSID starts with $\Gamma = 0$ since the IDS does not have yet any information about the network traffic patterns at the initial learning phase. That is, the output of the IDS is calculated using the initial parameter values (if available) and will not be able to achieve accurate detection for the particular traffic. 
In addition, for selecting the first packet, the parameters of the IDS are updated for the first time using $p_i^- = 1,~ p_i^+ = 0$, and $q_i = 0$. Thus, SSID selects the first packet to learn as a benign sample. 


\subsection{Trustworthiness of IDS}\label{sec:trust}

Now, we determine the trust coefficient $\Gamma$ for the IDS in the SSID framework. Through this coefficient, we aim to include both the effects of changes in the normal behavior of network traffic over time and the generalization ability of the IDS into the packet selection model for learning.

To this end, we first define the factor of ``representativeness'', denoted by $C_{rep}$, for the traffic packets that are learned by the IDS. The representativeness factor $C_{rep}$ takes a value in the range of $[0, 1]$ and measures how much the packets used for learning (during all of the past learning phases) represent the total observed traffic. 

In addition, we define the factor of ``generalization ability'', denoted by $C_{gen}$, of the IDS. The generalization factor $C_{gen}$ takes a value in the range of $[0, 1]$ and is calculated only at the end of each parameter update since it is the only time when the parameters of the IDS are updated. These two factors shall respectively be given in Section~\ref{sec:representativeness} and Section~\ref{sec:generalization}.

{Accordingly, in (}\ref{eq:trust}{), }
{we determine the trust coefficient $\Gamma$ }
{as the multiplication of $C_{rep}$ and $C_{gen}$:}
\begin{equation} \label{eq:trust}
	\Gamma =  C_{rep} \, C_{gen}
\end{equation}
In this way, $\Gamma$ simultaneously measures how much the IDS is able to learn and generalize from provided traffic packets  and how much these packets reflect actual traffic patterns. That is, through this trust coefficient, we evaluate how much information the IDS can generalize from the traffic packets provided to make decisions for the upcoming traffic. 

\subsection{Representativeness of the Traffic that is Learned ($C_{rep}$)} \label{sec:representativeness}

In order to calculate the representativeness of the packet traffic used during the earlier learning phases, we compare the learned traffic with the total observed traffic through Kullback-Leibler (KL)-Divergence \cite{KL}. Therefore, there are two sets of traffic packets for comparison, the packets used in the previous learning phases up to and including $l$ (where $l$ is the latest completed learning phase) and the normal packets that are observed by IDS during continuous detection.

During this comparison, we assume that the packet traffic consists of two main properties, inter-transmission time ($TT$) and the packet length ($PL$) since these properties can be considered as the basis of traffic metrics, which are the inputs of the IDS. {Inter-transmission time ($TT$) refers to the duration between the transmissions of consecutive packets, while packet length ($PL$) denotes the size of each packet in the network traffic. These properties are crucial as they directly reflect the behaviour and characteristics of network traffic, thus serving as fundamental metrics for IDS to analyse and detect anomalies or malicious activities effectively.} 

We further assume that packet arrivals --any sample collected from the network traffic-- has a Poisson distribution so that the inter-transmission time $TT$ is an Exponentially-distributed random variable. The packet length $PL$ is also assumed to be an Exponentially-distributed random variable because the header length is considerably larger than the message length for the majority of IoT applications. In addition, $TT$ and $PL$ are considered to be independent. On the other hand, for particular applications, these assumptions and the traffic model can be changed and the below methodology can easily be adapted for the new traffic model with a new set of assumptions.    

Furthermore, let $S^{TT}_l$ and $S^{PL}_l$ respectively denote the sets of the inter-transmission times and lengths of all packets learned at the end of $l$, and $S^{TT}_o$ and $S^{PL}_o$ respectively denote the same of all normal packets observed during continuous detection. In addition, according to our assumptions, $S^{TT}_l$ and $S^{TT}_o$ have exponential distributions with means of $1/\lambda_l$ and $1/\lambda_o$ while $S^{PL}_l$ and $S^{PL}_o$ also have exponential distributions with means of $1/\mu_l$ and $1/\mu_o$.  

\subsubsection{KL-Divergence for Inter-Transmission Times} \label{sec:divergence_time}
For the set of inter-transmission times, $D_{KL}(S^{TT}_o || S^{TT}_l)$ is KL-Divergence from $S^{TT}_l$ to $S^{TT}_o$ measuring the information gain achieved if $S^{TT}_o$ would be used instead of $S^{TT}_l$ which has been used during the learning phases of SSID. Note that small KL-Divergence means low information gain, and $D_{KL}(S^{TT}_o || S^{TT}_l) = 0$ shows that $S^{TT}_o$ and  $S^{TT}_l$ provide the same amount of information. Accordingly, using the definition of KL-Divergence \cite{KL}, we first calculate $D_{KL}(S^{TT}_o || S^{TT}_l)$, which can shortly be denoted by $D_{KL}^{TT}$, as 
\begin{eqnarray}\label{eqn:kl_transmissions_init}
	D_{KL}^{TT} &=& \int_{-\infty}^{\infty} f(x; \lambda_o) log(\frac{f(x; \lambda_o)}{f(x; \lambda_l)}) \,dx \\
	&=& \E_{f(x; \lambda_o)}\big[\, log(\frac{f(x; \lambda_o)}{f(x; \lambda_l)}) \, \big] \nonumber\\
	&=& \E_{f(x; \lambda_o)}\big[\, log(\frac{\lambda_o}{\lambda_l}) - x(\lambda_o - \lambda_l) \, \big]\nonumber
\end{eqnarray}
where $f(x; \lambda_o)$ and $f(x; \lambda_l)$ denote the probability distribution functions of $S^{TT}_o$ and $S^{TT}_l$ respectively with parameters $\lambda_o$ and $\lambda_l$. This leads to the result of 
\begin{equation}\label{eqn:kl_transmissions}
	D_{KL}^{TT} = log(\frac{\lambda_o}{\lambda_l}) - \frac{(\lambda_o - \lambda_l)}{\lambda_o}
\end{equation}

\subsubsection{KL-Divergence for Packet Lengths} \label{sec:divergence_lengths}
Similarly with transmission times, for the set of packet lengths, $D_{KL}(S^{PL}_o || S^{PL}_l)$ is KL-Divergence from $S^{PL}_l$ to $S^{PL}_o$, which is shortly denoted by $D_{KL}^{PL}$, and is calculated as
\begin{eqnarray}\label{eqn:kl_lengths_init}
	D_{KL}^{PL} &=&  \int_{-\infty}^{\infty} f(x; \mu_o) log(\frac{f(x; \mu_o)}{f(x; \mu_l)}) \,dx \\
	&=& \E_{f(x; \mu_o)}\big[\, log(\frac{f(x; \mu_o)}{f(x; \mu_l)}) \, \big] \nonumber\\
	&=& \E_{f(x; \mu_o)}\big[\, log(\frac{\mu_o}{\mu_l}) - x(\mu_o - \mu_l) \, \big]\nonumber
\end{eqnarray}
where $f(x; \mu_o)$ and $f(x; \mu_l)$ denote the probability distribution functions of $S^{PL}_o$ and $S^{PL}_l$ respectively with parameters $\mu_o$ and $\mu_l$. This results in: 
\begin{equation}\label{eqn:kl_lengths}
	D_{KL}^{PL} = log(\frac{\mu_o}{\mu_l}) - \frac{(\mu_o - \mu_l)}{\mu_o}
\end{equation}

\subsubsection{Representativeness Factor based on Normalized KL-Divergence}
For both transmission times and packet lengths, we now obtained the KL-Divergence between the set of observed packets and the set of packets learned. 
However, the KL-Divergence cannot directly be used as a representativeness factor because of the following reasons: 1) It has no upper bound but the representativeness factor $C_{rep} \in [0, 1]$. 2) KL-Divergence is decreasing function of the similarity between two sets but we need an increasing function of that as the name ``representativeness'' suggests. 3) This factor should be the combination of $D_{KL}^{TT}$ and $D_{KL}^{PL}$. 

Therefore, in order to obtain the representativeness factor, we first normalize each of $D_{KL}^{TT}$ and $D_{KL}^{PL}$ as 
\begin{eqnarray}
	&D_{KL-norm}^{TT} = e^{-D^{TT}_{KL}},\\
	&D_{KL-norm}^{PL} = e^{-D^{PL}_{KL}}.
\end{eqnarray}
which solve the issues 1) and 2) stated above. Each of these normalized divergence measures can also be written in terms of only the traffic parameters:
\begin{eqnarray}\label{eqn:KL_normalized_TT}
	D_{KL-norm}^{PL} &=&  e^{-\big[\, log(\frac{\lambda_o}{\lambda_l}) - \frac{(\lambda_o - \lambda_l)}{\lambda_o} \, \big]} \nonumber\\
	&\nonumber\\ 
	&=& \Bigl[\, \frac{\lambda_l}{\lambda_o} \,  \,  e^{-\frac{(\lambda_l - \lambda_o)}{\lambda_o}} \, \Bigr]
\end{eqnarray}
Similarly, 
\begin{eqnarray}\label{eqn:KL_normalized_PL}
	D_{KL-norm}^{PL} &=&  e^{-\big[\, log(\frac{\mu_o}{\mu_l}) - \frac{(\mu_o - \mu_l)}{\mu_o} \, \big]} \nonumber\\
	&\nonumber\\ 
	&=& \Bigl[\, \frac{\mu_l}{\mu_o} \,  \,  e^{-\frac{(\mu_l - \mu_o)}{\mu_o}} \, \Bigr]
\end{eqnarray}

Then, we combine $D_{KL-norm}^{TT}$ and $D_{KL-norm}^{PL}$ into the ``representativeness factor'' $C_{rep}$ as 
\begin{equation}
	C_{rep} = c_1 D_{KL-norm}^{TT} + c_2 D_{KL-norm}^{PL}  
\end{equation}
where $c_1 \leq 1$ and $c_2 \leq 1$ are positive constants that satisfy $c_1 + c_2 = 1$. 

In order to weigh transmission times and packet lengths equally, we take $c_1 = c_2 = 0.5$. That is, we take their average:
\begin{eqnarray}\label{eq:c_rep_DL}
	C_{rep} &=& \frac{1}{2} \big[D_{KL-norm}^{TT} + D_{KL-norm}^{PL}\big] \\ 
	&\nonumber\\ 
	&=& \frac{1}{2} \big[e^{-D^{TT}_{KL}} + e^{-D^{PL}_{KL}}\big]  \nonumber
\end{eqnarray}
We can rewrite {(}\ref{eq:c_rep_DL}{)} only in terms of the traffic parameters using (\ref{eqn:KL_normalized_TT}) and (\ref{eqn:KL_normalized_PL}):
\begin{equation} \label{eq:rep_factor}
	C_{rep} = \frac{1}{2} \Bigl[\, \frac{\lambda_l}{\lambda_o} \,  \,  e^{-\frac{(\lambda_l - \lambda_o)}{\lambda_o}} + \frac{\mu_l}{\mu_o} \,  \,  e^{-\frac{(\mu_l - \mu_o)}{\mu_o}} \, \Bigr]
\end{equation}


\subsection{Generalization Ability of IDS ($C_{gen}$)} \label{sec:generalization}
As stated above, we consider the generalization ability of the IDS as one of two factors that define the trustworthiness of intrusion decisions. To this end, the aim of this subsection is to determine the generalization ability of the IDS in simple terms to make its computation as easy as possible using the available measures during the execution of SSID. Accordingly, we start with the basic definition of generalization \cite{statistical_learning_theory}: 
\begin{center}
	Generalization $\equiv$ Data + Knowledge
\end{center}
stating that the generalization depends on the ``Data'', which is denoted by $\Delta$ and refers to the adequacy of the packet samples that are used for learning, and the ``Knowledge'', which is denoted by $\kappa$ and refers to the knowledge of the IDS obtained from packets learned. Therefore, we define the generalization factor $C_{gen}$ as
\begin{equation} \label{eq:def_gen_factor}
	C_{gen} = c_3 \, \Delta + c_4 \, \kappa
\end{equation}
where $c_3$ and $c_4$ are positive constants such that $c_3, c_4 \leq 1$, and $c_3 + c_4 = 1$. 

\subsubsection{Data Adequacy ($\Delta$)} We evaluate the adequacy of the packet samples that are used for learning with respect to the number of learnable parameters in the IDS. Although there is no hard rule for determining the adequacy of the learning data (i.e., the number of training samples required) for a given DL model, most studies have shown its relationship to the total number of learnable parameters in the model and taken the minimum number of required training samples as a multiple of the number of parameters \cite{alwosheel2018_sample_size}. 

{Therefore, we define the $\Delta$ as the counterpart of the ratio of the number of learnable parameters in the IDS to the total number of packet samples used for learning up to and including learning phase $l$: }
\begin{equation}\label{eq:data_adequacy}
	\Delta = \Bigl[1 - \min\Bigl(\frac{W}{\sum_{k=0}^{l}{|B^k|}}, 1\Bigr)\Bigr]
\end{equation}
{where $\sum_{k=0}^{l}{|B^k|}$ is the total number of packet samples that are sequentially used to learn model parameters until the end of learning phase $l$. Clearly, $\Delta$ takes value in $[0, 1]$. While $W$ is a constant number and $|B^l| \geq 1$ for any learning phase $l$ (in which a learning is performed), $\lim_{l \to\infty}(\Delta) = 1$. In addition, $\Delta = 0$ when $\sum_{k=0}^{l}{|B^k|} \leq W$. }



\subsubsection{Knowledge ($\kappa$)}
We consider knowledge to be the measure of the DL model's expected performance for the upcoming traffic packets. Subsequently, we measure the knowledge (i.e. expected performance) of the IDS based on its performance on the packet samples used for learning and on the online available validation data.


To this end, in this paper, we consider the worst-case scenario when there is no validation data available. $\mathcal{E}(l)$ is the empirical error measured at the end of learning phase $l$ on both packet samples learned and validation data (if available), such that $0 \leq \mathcal{E}(l) \leq 1$. We then define the knowledge ${\kappa}$ as the counterpart of the exponentially weighted moving average of empirical errors for all learning phases up to and including the $l$--th phase:
\begin{equation}\label{eq:knowledge}
	\kappa = 1 - \sum_{k=0}^{l}{(\frac{1}{2})^{(l-k+1)} \, \mathcal{E}(k)}
\end{equation}
where the multiplier is set as $1/2$ to keep the value of $\kappa$ in $[0, 1]$. That is, if the empirical training error decreases with the successive learning phases (i.e. $\mathcal{E}(l)$ is the decreasing function of $l$), the knowledge of the IDS increases converging to its maximum.

In practice, at the end of each learning phase $l$, $\kappa$ can easily be updated using only its previous value and the empirical error $\mathcal{E}(l)$ as 
\begin{equation}\label{eq:knowledge_update}
	\kappa \leftarrow \frac{1}{2} - \frac{1}{2}\Bigr[ \mathcal{E}(l) - \kappa \Bigl]
\end{equation}

\subsubsection{Generalization Factor}

We now easily calculate the generalization factor $C_{gen}$ combining the ``data adequacy'' $\Delta$ (\ref{eq:data_adequacy}) and ``knowledge'' $\kappa$ (\ref{eq:knowledge}) using the definition of the generalization factor (\ref{eq:def_gen_factor}): 
\begin{align}
	&C_{gen} =   \\
	& c_3 \Bigl[1 - \min\bigl(\frac{W}{\sum_{k=0}^{l}{|B^k|}}, 1\bigr)\Bigr] +  c_4 \Bigl[ 1 - \sum_{k=0}^{l}{(\frac{1}{2})^{(l-k+1)} \, \mathcal{E}(k)} \Bigr]  \nonumber
\end{align}

We particularly set $c_3 = c_4 = 0.5$ representing that the data and knowledge are equally important for generalization:
\begin{align}\label{eq:gen_factor}
	&C_{gen} = \\
	&1 -  \frac{\min\bigl(W/\sum_{k=0}^{l}{|B^k|}, 1\bigr) + \sum_{k=0}^{l}{(1/2)^{(l-k+1)} \, \mathcal{E}(k)}}{2} \nonumber
\end{align}


\section{Results}\label{sec:Results}

We now evaluate the performance of SSID framework for two different intrusion detection tasks to identify malicious traffic packets and compromised devices. 


\subsection{Parameter Settings for SSID}

{First, we set the parameters of SSID as follows: trust threshold $\Theta = 0.95$, number of packets observed for decision $I = 10$, minimum number of training packets $K = 100$, and intrusion threshold $\gamma=0.25$. That is, SSID aims to keep the trust in the IDS above $0.95$ while it considers a packet as malicious if the output of the IDS is above $0.25$. IDS analyses $I = 10$ packets to make robust intrusion decisions, which allows IDS to make early decisions while not being too reactive to instantaneous changes. In addition, the parameters are updated using at least $K = 100$ packets in a learning batch for computational efficiency. 
}

\subsection{{Datasets}}
{Since the proposed SSID framework provides online learning for the IDS, its performance is evaluated using two well-known datasets, \textbf{Kitsune} }\cite{kitsune_paper, kitsune_dataset}{ and \textbf{Bot-IoT} }\cite{botiot_dataset}{, which contain the actual packet transmissions for both normal and malicious traffic over time. These are two of the most recent and used datasets on Botnet and DDoS attacks. }

{Some other examples of such datasets are UNSW-NB15 }\cite{UNSW_NB15_dataset}{, CICIDS 2017 }\cite{CCIDS2017_dataset}{, and IoT-23 }\cite{iot23dataset}{. The UNSW-NB15 dataset focuses on general network intrusion detection, not specifically focuses on botnet activities, capturing a wide range of network activities, while CICIDS 2017 emphasizes both traditional and IoT-specific attacks in controlled lab environments. IoT-23 is tailored for studying IoT-specific security challenges, offering data from various IoT devices. Meanwhile, Bot-IoT and Kitsune datasets specifically focuses on IoT DoS and DDoS attacks. They include realistic IoT device behaviors and various types of botnet activities, such as command and control communication, malware propagation, and reconnaissance scans. These datasets offer a comprehensive collection of normal and malicious traffic data for benchmarking attack detection and compromised device identification systems.}

{In the rest of this section, for malicious traffic detection and compromised device identification during Mirai Botnet attack, we first use the well-known Kitsune dataset} \cite{kitsune_dataset}{, which contains $764,137$ packet transmissions of both normal and attack traffic cover a consecutive time period of roughly $7137$ seconds. Of the total $764,137$ packets exchanged between $107$ distinct IP addresses in this dataset, $121,621$ are normal packets and $642,516$ are malicious packets. Then, for compromised device identification, in addition to the Mirai Botnet, we use the following data: 1) SYN DoS attack from the Kitsune dataset, 2) DDoS attacks using HTTP, TCP and UDP protocols from the Bot-IoT dataset, and 3) DoS attack using HTTP protocol from the Bot-IoT dataset. The data of SYN DoS, DDoS HTTP, DDoS TCP, DDoS UDP, and DoS HTTP are respectively comprised of ``$2,771,276$'', ``$19,826$'', ``$19,548,235$'', ``$18,965,736$'', and ``$29,762$'' packets. 
}

\subsection{Performance Evaluation for Malicious Traffic Detection}

\begin{figure}[h!]
	\centering
	\includegraphics[scale=0.24]{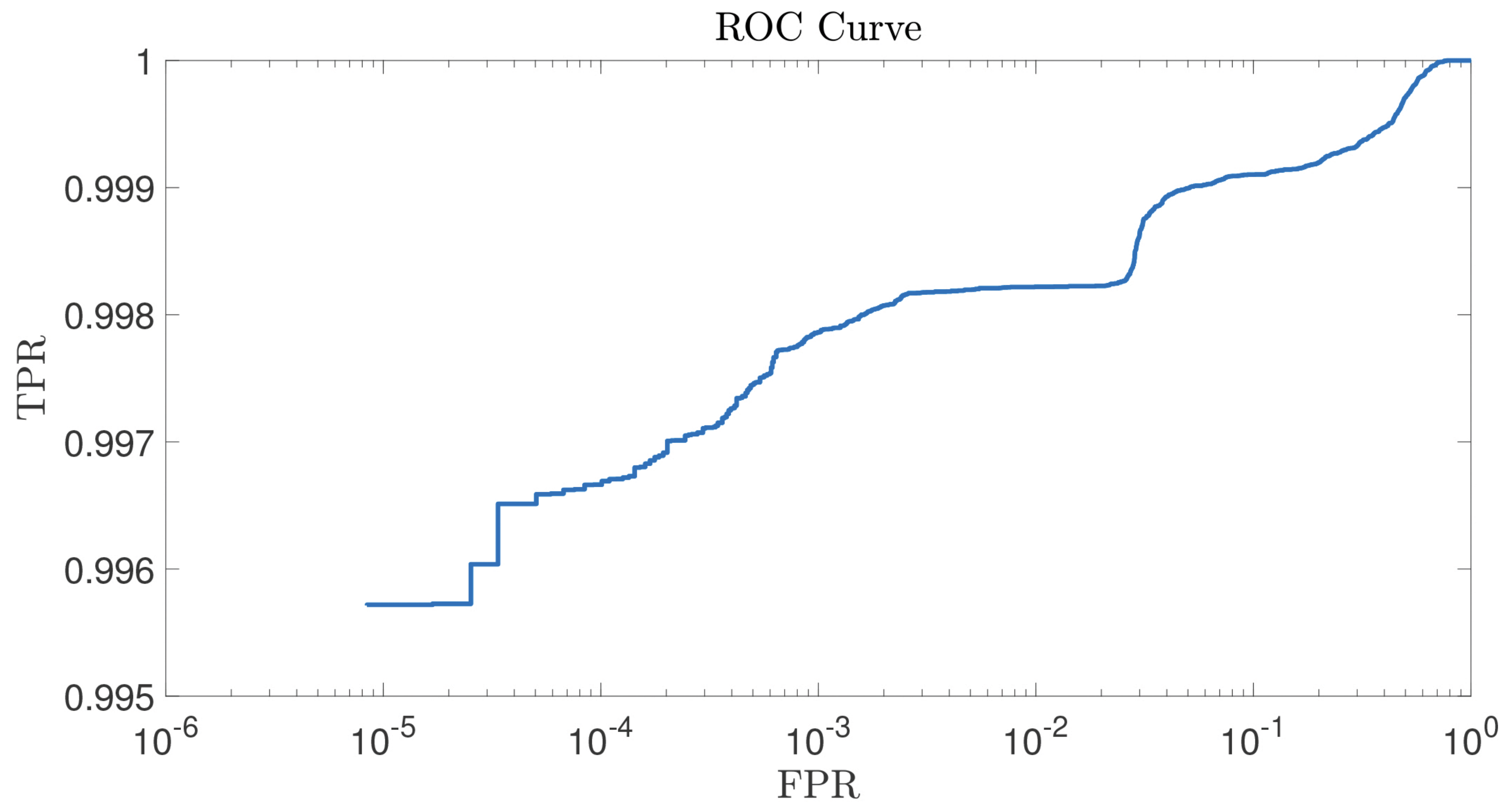}
	\caption{ROC curve for the performance of AADRNN-based IDS under the SSID framework for malicious traffic detection}
	\label{fig:SSID_malicious_ROC}
\end{figure}

The performance of SSID is first evaluated for malicious traffic detection during Mirai Botnet attack. 
Figure~\ref{fig:SSID_malicious_ROC} displays the ROC curve, where the x-axis of this figure is plotted in logarithmic scale. We see that AADRNN-based IDS trained under our novel SSID framework achieves significantly high TPR above $0.995$ even for very low FPR about $10^{-5}$. 

\begin{figure}[h!]
	\centering
	\includegraphics[scale=0.25]{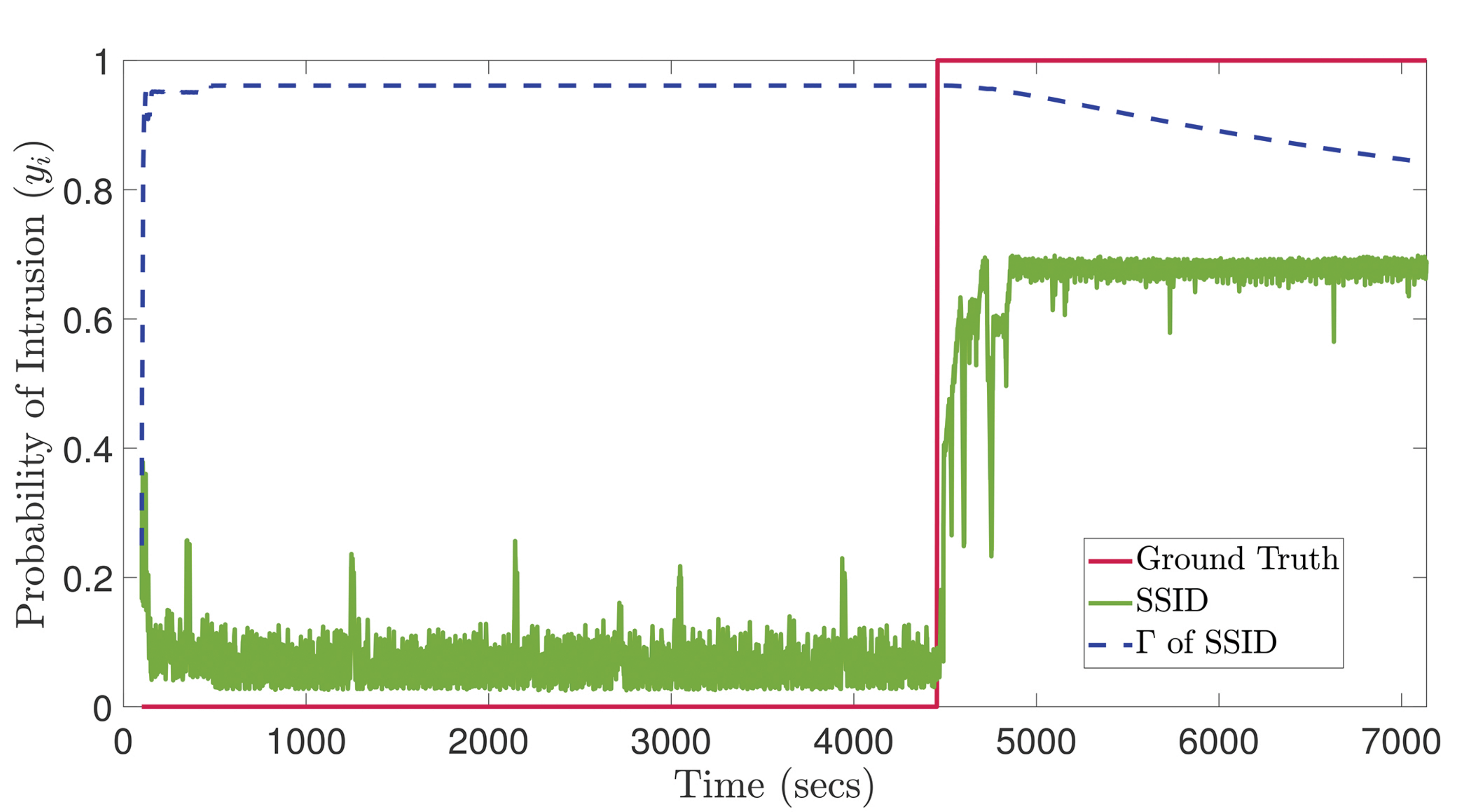}
	\caption{Predictions of SSID and the value of trust coefficient $\Gamma$ with respect to time 
	}
	\label{fig:SSID_malicious_predictions}
\end{figure}

In more detail, in Figure~\ref{fig:SSID_malicious_predictions}, we present the predictions and $\Gamma$ of SSID with respect to time. This figure reveals an important fact that while the IDS is completely indecisive at the beginning, SSID framework enables it to learn the normal traffic very quickly. As a result, SSID makes significantly low false alarms although it learns --fully online-- during real-time operation based only on its own decision using no external (offline collected) dataset. We also see that $\Gamma$ accurately reflects the trustworthiness of decisions made by AADRNN. In addition, although $\Gamma$ slightly decreases as a result of random packet selection, especially after attack starts, the parameters of AADRNN are not updated by SSID as the traffic is detected as malicious.      

\subsubsection{Comparison with Incremental and Offline Learning}

We further compare the performance of AADRNN under SSID with the performance of AADRNN with incremental and offline learning. All methods with offline learning are trained using the first $83,000$ benign traffic packets while the AADRNN with incremental learning is trained periodically for the window of $750$ packets using AADRNN's own decision,  where the first $750$ packets received are assumed to be normal packets during the cold-start of the network. 

\begin{figure}[h!]
	\centering
	\includegraphics[scale=0.25]{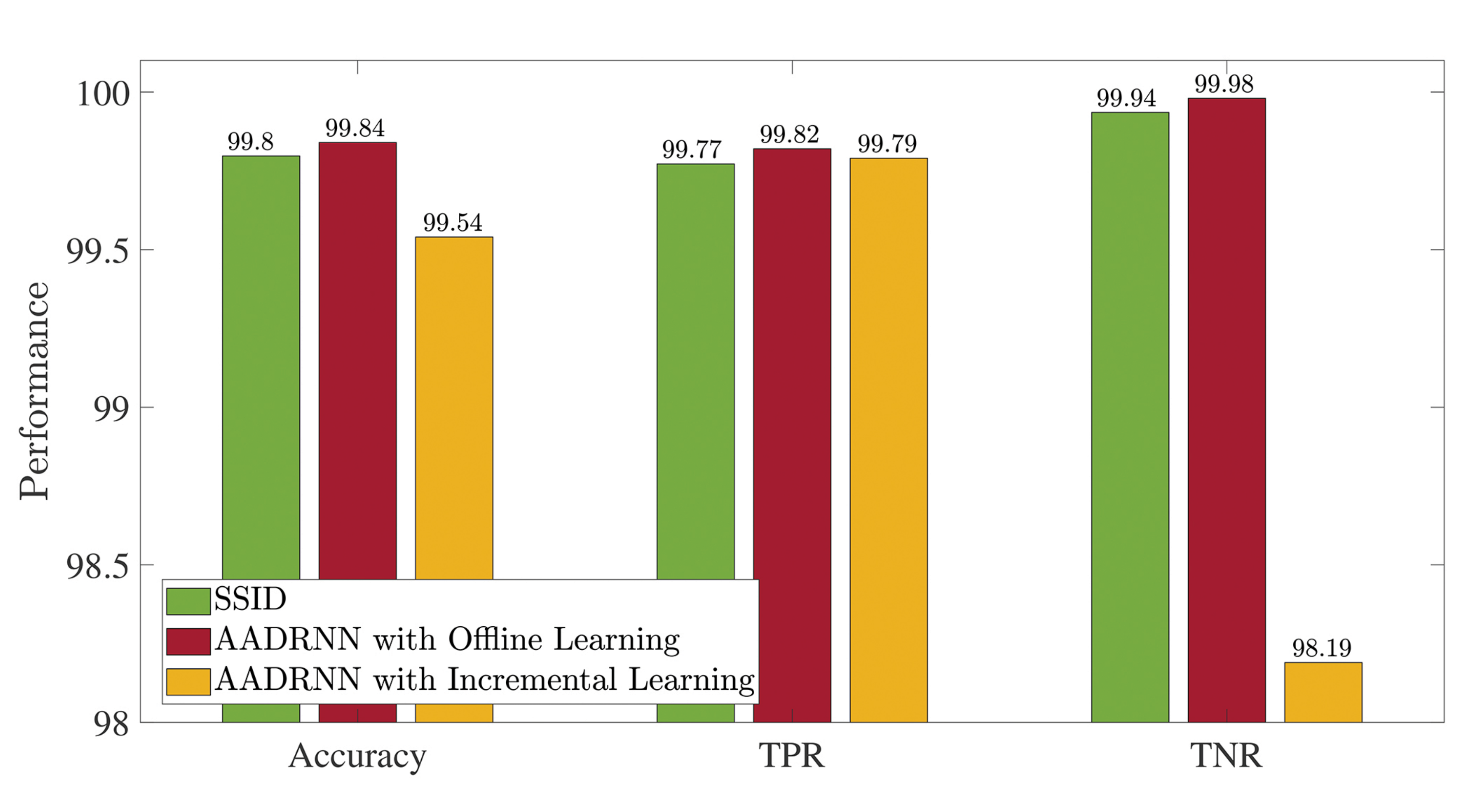}
	\caption{Performance comparison between the AADRNN under SSID and the AADRNN with incremental and offline learning}
	\label{fig:SSID_AADRNN_comparison}
\end{figure}

Figure~\ref{fig:SSID_AADRNN_comparison} displays the performances of SSID and the AADRNN,  with incremental and offline learning. The results in this figure first reveal that the fully online trained AADRNN using the SSID framework achieves competitive results with the AADRNN which is trained offline using approximately $83,000$ packets. We also see that the SSID significantly outperforms the AADRNN with incremental learning with respect to all performance metrics. 
Also note that the SSID learned from a total of $4,161$ packets while also conducting  real-time attack detection.

In contrast with offline and incremental learning, SSID framework assumes only that the first packet is known to be benign so the duration of cold-start equals the transmission of a single traffic packet. That is, using no offline dataset or requiring no cold-start, the SSID framework is able train an DL-based IDS to achieve considerably high performance which is highly competitive against the DL models trained on significantly large dataset.  

\subsubsection{A Different DL Model --MLP-- under the SSID Framework}

During the performance evaluation of the SSID framework, we also use MLP, which is one of the most popular feed-forward neural networks used for various tasks such as signal processing, forecasting, anomaly detection, etc. As also reviewed in Section~\ref{sec:RelatedWorks}, various works \cite{Tuan, Doshi, Letteri, Soe} used MLP to develop different IDS methods. 

Similar to the AADRNN, the MLP model that we use is also comprised of $M$ layers with $M$ neurons each. Each neuron has \emph{sigmoid} activation function as
\begin{equation}
	\zeta(\Lambda) = \frac{1}{1 + e^{-\Lambda}},
\end{equation}
where $\Lambda$ is an input to the neural activation. 

In both the initial and online learning stages, the parameters of the MLP are updated using the state-of-the-art optimizer Adam. 
In each online learning phase, incremental learning is applied by starting parameter optimization from the connection weight values already in use at the beginning of that phase.

\begin{figure}[h!]
	\centering
	\includegraphics[scale=0.25]{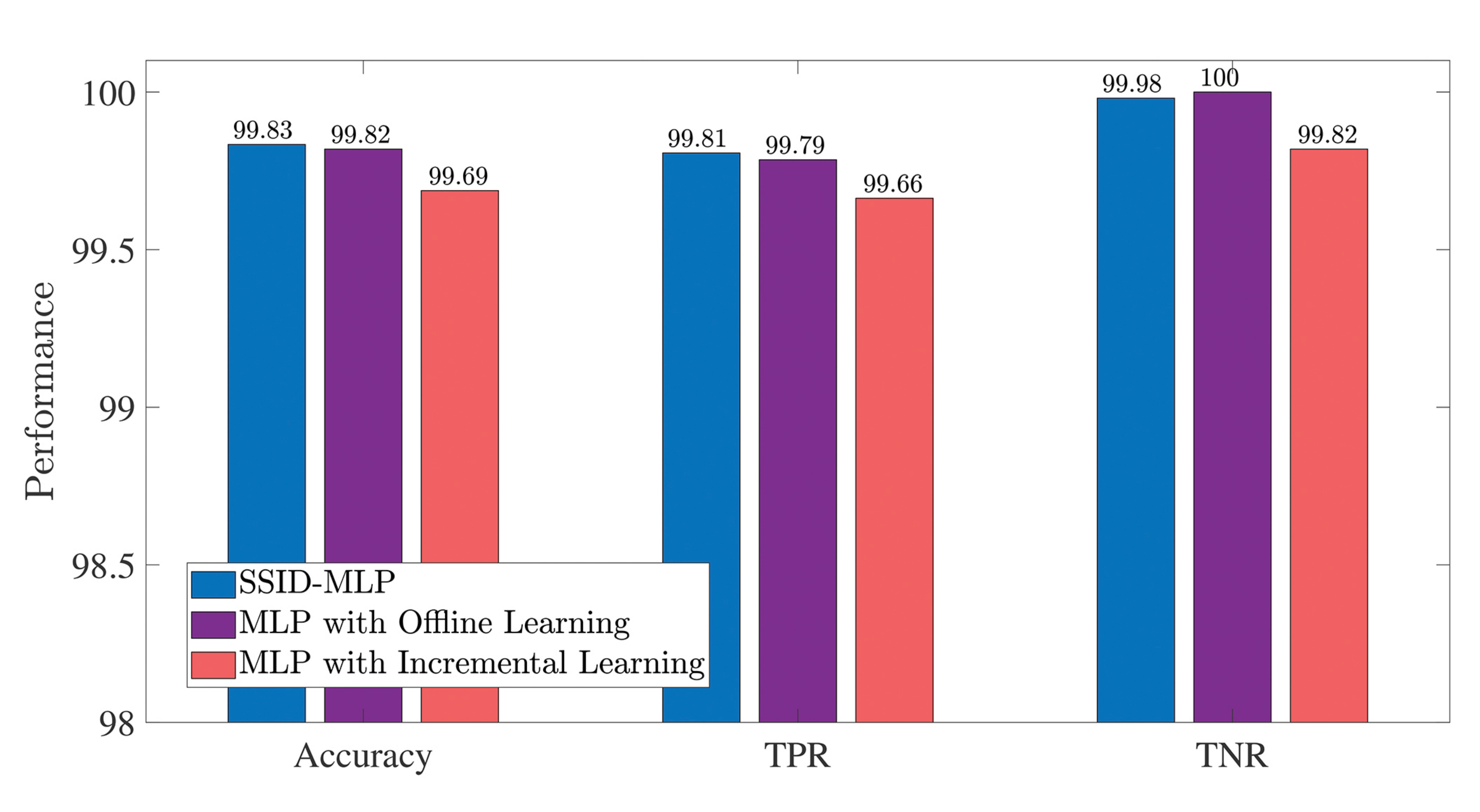}
	\caption{Performance comparison between the SSID-MLP and the MLP with incremental and offline learning}
	\label{fig:SSID_MLP_comparison}
\end{figure}

In order to further analyze the impact of the proposed SSID framework on the performance of a different DL model, we evaluate the performance of the well-known MLP under the SSID framework (called SSID-MLP) and compare it with the performance of MLP with offline and incremental learning, respectively. The results of this performance evaluation is presented in Figure~\ref{fig:SSID_MLP_comparison}.

Figure~\ref{fig:SSID_MLP_comparison} shows that SSID-MLP achieves slightly higher Accuracy and TPR than MLP with offline learning, although MLP with offline learning raises no false alarms at all (i.e. with $100 \%$ TNR). Moreover, we see that SSID-MLP significantly outperforms the MLP model that is also trained via incremental learning periodically for every $750$ packets based on its own output.

\subsubsection{Comparison of Different DL Models}

\begin{figure}[h!]
	\centering
	\includegraphics[scale=0.25]{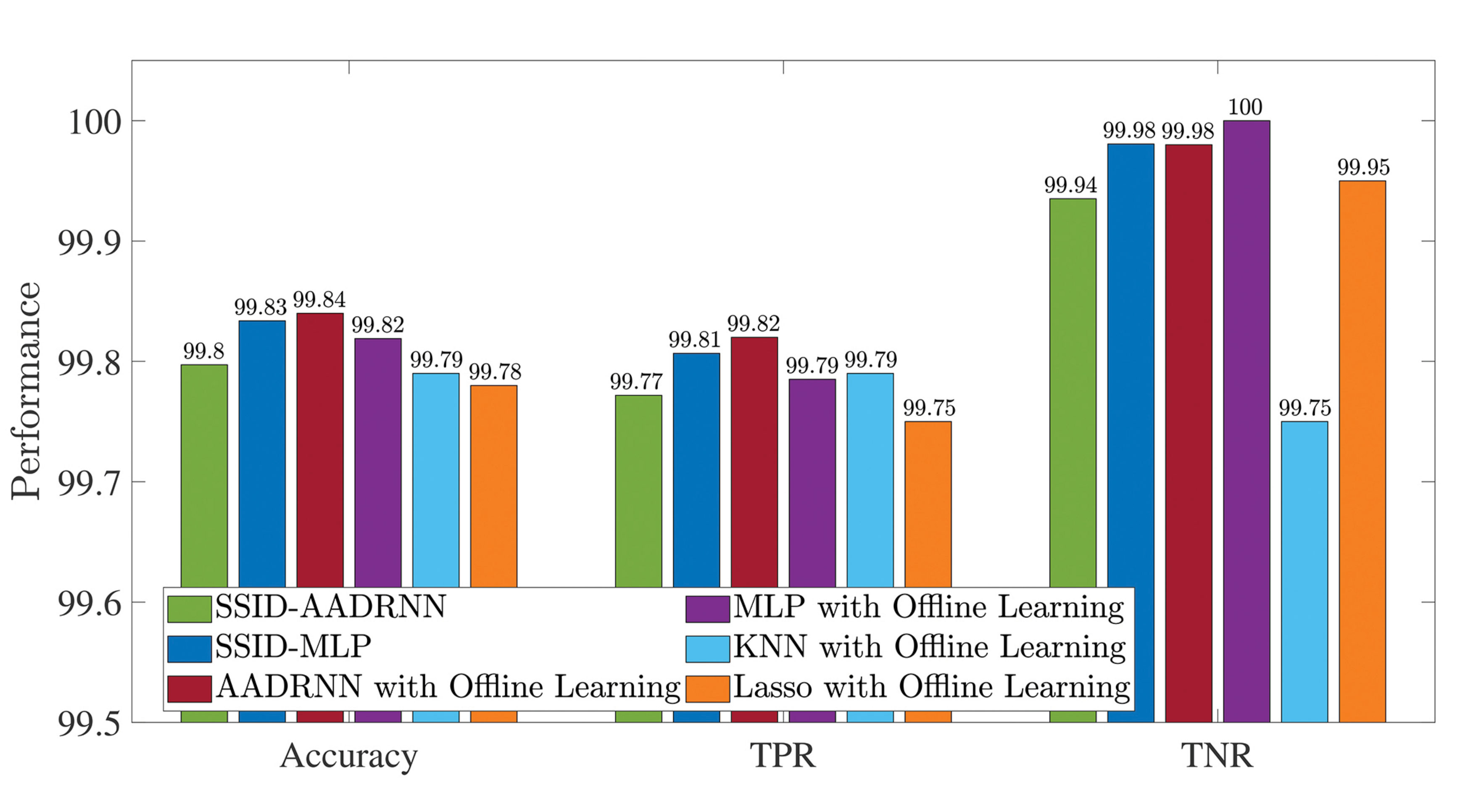}
	\caption{Performance comparison between the ML models under the SSID framework and those with offline learning}
	\label{fig:SSID_comparison}
\end{figure}

{We further compare the performance of AADRNN under SSID (called SSID-AADRNN for clarity) and SSID-MLP with those of some well-known ML models, including KNN and Lasso with offline learning. Note that models with offline learning are trained using a labeled normal traffic data. Also, note that Lasso, KNN, and MLP are well-known and commonly used models for Botnet attack detection and compromised device identification }\cite{Doshi, Letteri, Sriram, Nguyen_Botnet_IIoT}.  Figure~\ref{fig:SSID_comparison} displays the performance of all compared models with respect to Accuracy, TPR and TNR. The results in this figure show that SSID-MLP achieves the second-best performance with respect to all performance metrics. In addition, both SSID-MLP and SSID-AADRNN achieve highly competitive results with the offline trained DL models, while the SSID framework completely eliminates the need for data collection and labeling.

\subsection{Performance Evaluation for Compromised Device Identification}

We now evaluate the performance of CDIS \cite{CDIS} (using the methodology in Section~\ref{sec:IDS}) under the SSID framework, in short SSID-CDIS, on six different attacks, from the two distinct datasets Kitsune and Bot-IoT. For each dataset, the performance of SSID-CDIS is compared with the CDIS technique with sequential learning. Using the same methods as in \cite{CDIS}, compromised device identification is performed for a $10$ seconds long time window. The performance is evaluated using the Balanced Accuracy \cite{balanced_accuracy}.

\begin{figure}[h!]
	\centering
	\includegraphics[scale=0.25]{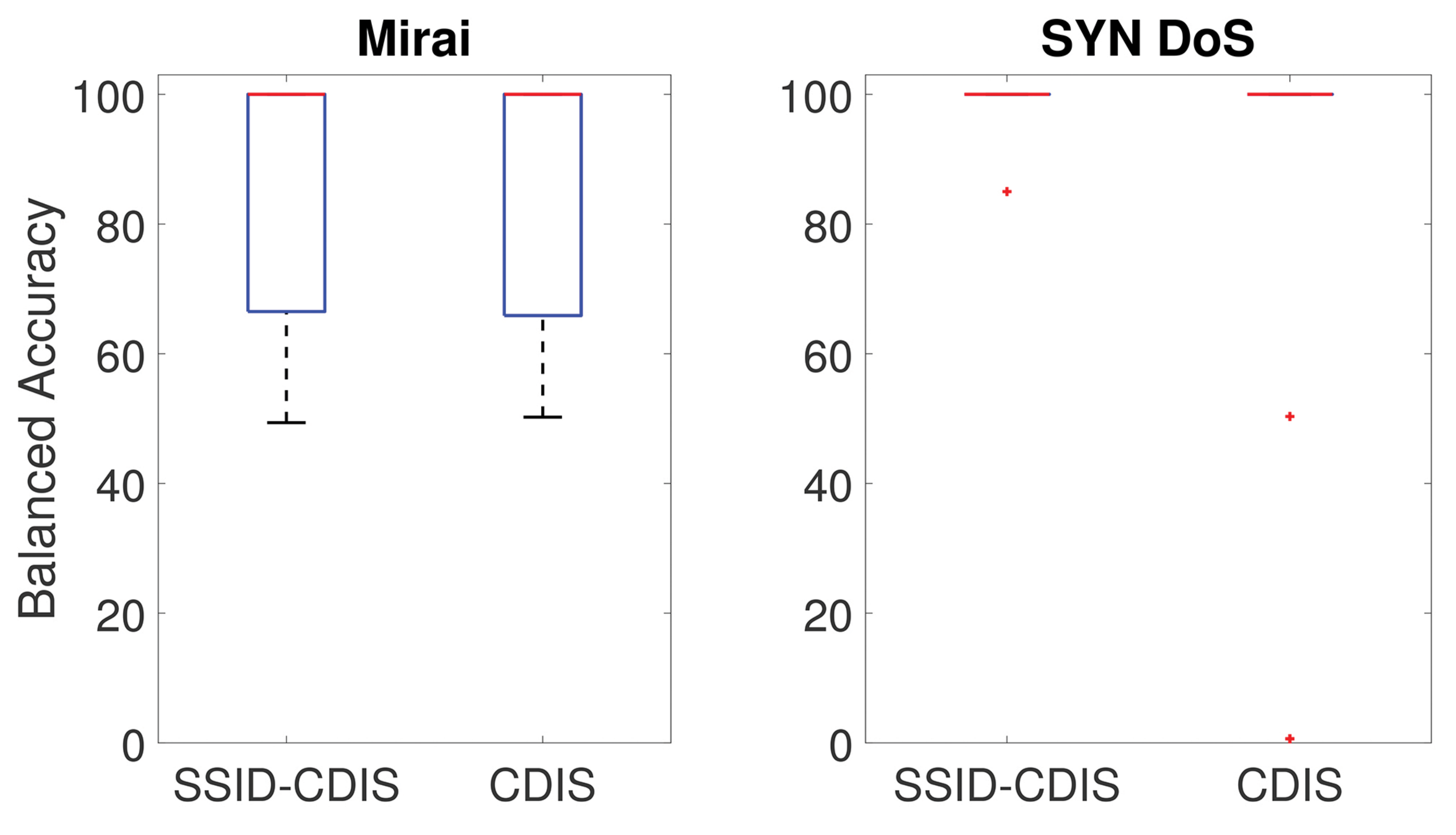}
	\caption{Performance comparison of the CDIS trained under the SSID framework with that under sequential learning on Kitsune dataset}
	\label{fig:SSID_compromised_Kitsune}
\end{figure}

Figure~\ref{fig:SSID_compromised_Kitsune} displays the performance of SSID-CDIS and its comparison with CDIS to identify compromised IP addresses, for each of the Mirai Botnet and SYN DoS attacks in the Kitsune dataset. Specifically it shows that while the SSID framework provides the same performance as sequential learning to identify compromised devices during a Mirai Botnet attack, it significantly improves the overall performance of CDIS during a SYN DoS attack. The box plot on the right of Figure~\ref{fig:SSID_compromised_Kitsune} shows that SSID-CDIS achieves $100\%$ median balanced accuracy when there is only one outlier IP address with around $85\%$ accuracy. On the other hand, the sequentially trained CDIS has two outlier IP addresses with performances of $50\%$ and $1\%$, respectively.


\begin{table}[h!]
	\centering
	\normalsize
	\caption{{Comparison of Average Performance over IP Addresses between SSID-CDIS and Different ML Models for Mirai Botnet from the Kitsune Dataset}}
	\begin{tabular}{|c|| c | c|c|}
		\hline
		\begin{tabular}[c]{@{}c@{}}\textbf{Models}\end{tabular} & \begin{tabular}[c]{@{}c@{}}\textbf{Balanced} \\\textbf{Accuracy} \end{tabular} & \textbf{TPR} & \textbf{TNR} \\ \hline
		SSID-CDIS & 89.1 & 60.6 & 87.7 \\\hline
		CDIS & 87.7 & 90.3 & 79.4 \\\hline
		MLP  & 82.7 & 67.5 & 78 \\  \hline
		Lasso  & 85.1 & 86.7 & 75.6\\  \hline
		KNN  & 82.1 & 74.4 & 74.1  \\  \hline
	\end{tabular}\label{table:performance_compromised}
\end{table}

{In Table~}\ref{table:performance_compromised}{, we present the average performance of the SSID-CDIS against each of Lasso, KNN, and MLP with respect to Balanced Accuracy, TPR and TNR on Kitsune Mirai dataset. 
The results in this table show that SSID-CDIS achieves much higher Balanced Accuracy and TNR than other models. Although its average TPR is considerably low, SSID-CDIS provides a reasonable compromised device identification accuracy with a much lower false alarm rate compared to other models.}

\begin{figure}[h!]
	\centering
	\includegraphics[scale=0.25]{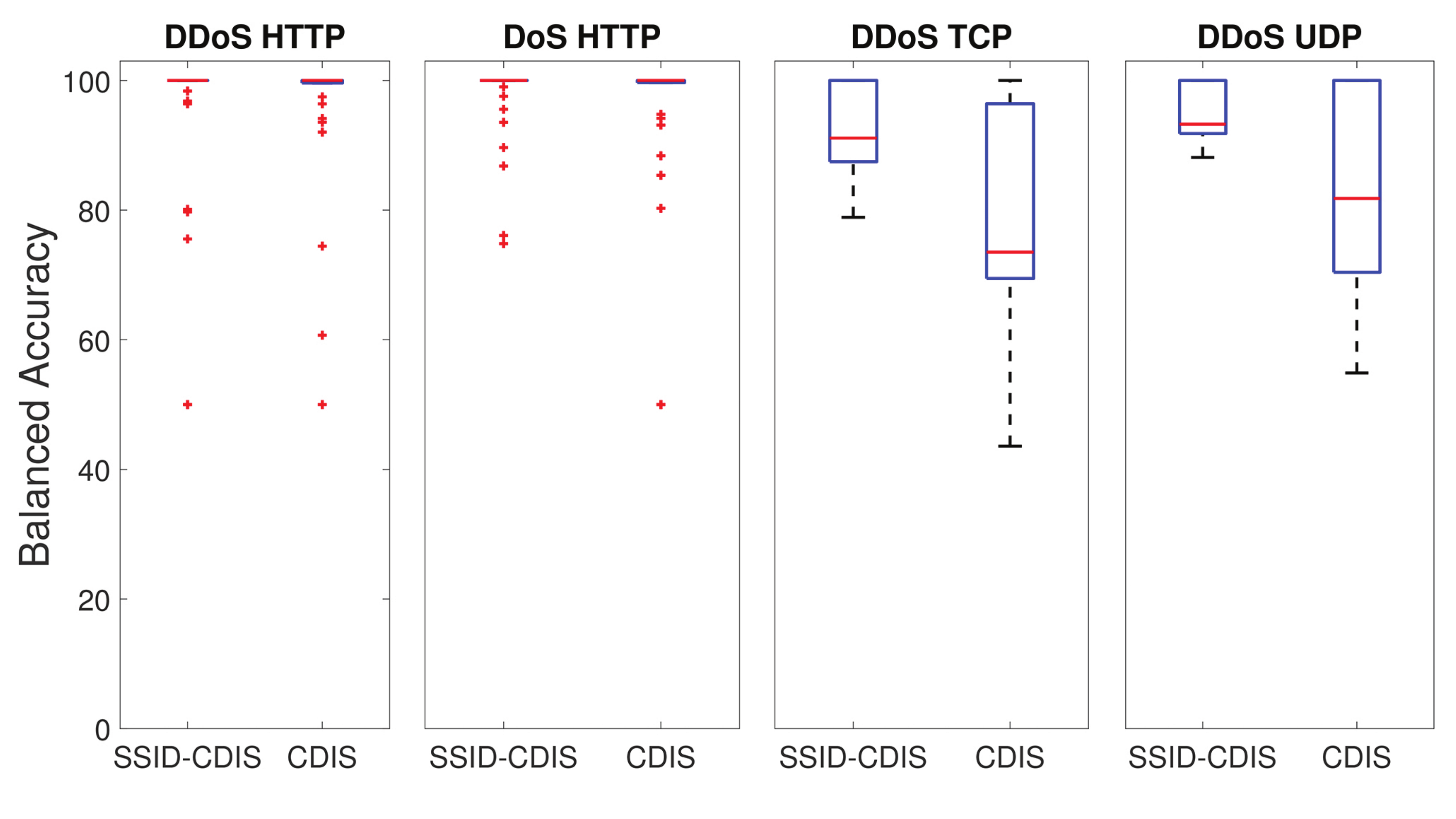}
	\caption{Performance comparison of the CDIS trained under the SSID framework with that under sequential learning on Bot-IoT dataset}
	\label{fig:SSID_compromised_BotIoT}
\end{figure}

Figure~\ref{fig:SSID_compromised_BotIoT} exhibits the performance of the SSID-CDIS system, and compares it with CDIS, to identify compromised IP addresses during DDoS and DoS attacks, using different communication protocols available in the Bot-IoT dataset. These results show that the SSID framework achieves higher identification performance, as compared to the use of CDIS sequential learning for the majority of attack types. 

Starting with the box plot displayed at the far left of this figure, we observe the following results:
\begin{enumerate}
	\item For the DDoS HTTP attack, the overall performance is almost the same for SSID and CDIS with sequential learning. However, as expected, performance varies slightly for individual IP addresses.
	\item For the DoS HTTP attack, using SSID improved the performance by $2 \%$ on average with a minimum of $75 \%$ balanced accuracy. 
	\item For the DDoS TCP attack, SSID significantly improved the median accuracy by $18 \%$, where SSID-CDIS achieves $91 \%$ median accuracy. In addition, while the balance accuracy of CDIS with sequential learning is below $80 \%$ (with a minimum of $49 \%$) for $9$ out of $13$ unique IP addresses, the balance accuracy of SSID-CDIS is equal to $79 \%$ for only $2$ IP addresses and above $80 \%$ for the rest. 
	\item Similar to the results for the DDoS TCP attack, SSID was seen to provide significant performance improvement to identify compromised devices during a DDoS UDP attack. The median accuracy increased by $11 \%$, achieving above $88 \%$ balanced accuracy for all IP addresses.
	\end{enumerate}


\section{Conclusion}\label{sec:Conclusion}

This paper has proposed a novel Self-Supervised Intrusion Detection (SSID) framework which is designed to train any given IDS (whose parameters are calculated using the network traffic) {\bf fully online} with no need for human intervention or prior offline training. The SSID framework comprises two successive learning stages, namely  initial learning and online learning. Initial learning aims to quickly adapt the IDS parameters to the network where the IDS is deployed. Online learning aims to update the parameters whenever it is  required to ensure high detection accuracy of the IDS. 

During the real-time operation of the IDS, in parallel to detection, the SSID framework performs the following main tasks: 
\begin{itemize}
	\item It continually estimates the trustworthiness of intrusion decisions to identify normal and malicious traffic. It also measures the ability of the IDS to learn and generalize from data provided by SSID and the extent to which this data can represent the current online network traffic patterns. 
	\item In order to provide training data for the IDS, the SSID framework selects and labels network traffic packets in a self-supervised manner based only on the decisions of IDS, and on the trust of SSID with regard to those decisions. 
	\item The SSID framework determines when the IDS parameters need to be updated, based on the trustworthiness of the IDS, the selected training packets, and the latest state of network security. 
\end{itemize}  
Thus the proposed SSID framework eliminates the need for offline data collection, it prevents human errors in data labeling avoiding labor and computational costs for model training and data collection through experiments. Its most important advantage is in terms of performance, and it enables IDS to easily adapt to the time varying characteristics of the network traffic.

We also evaluated the performance of the SSID framework for two tasks: malicious traffic detection and compromised device identification to enhance the security of an IoT network. For malicious traffic detection, two different DL models, AADRNN and MLP, have been deployed with the SSID framework and tested on the Kitsune dataset. The results we obtain reveal that the DL models trained under the SSID framework without offline training also achieve considerably high performance compared to the same models with offline and incremental learning. 

For compromised device identification, the performance of the state-of-the-art CDIS has been tested under sequential learning and the SSID framework on data from six different cyberattacks provided by the two publicly available Kitsune and Bot-IoT datasets. The results show that the use of SSID significantly improves the performance of CDIS for the majority of cases considered. 

Future work will evaluate the use of SSID for adapting a pre-trained IDS for use across different networks 
whose traffic has not been learned {\em a priori}, which seems to be a promising approach for fast, self-supervised, and successful adaptation of the IDS parameters for various networks. It would also be interesting to examine security assurance methods targeting distributed systems that combine the SSID framework with Federated Learning and attack prevention or mitigation algorithms. It seems that a successful integration of the SSID framework with Federated Learning may provide secure, distributed and self-supervised online learning for collaborative systems.

\appendix \label{sec:Appendix}

Table~\ref{table:abbreviation_SSID} and Table~\ref{table:symbols_SSID} respectively display the list of abbreviations and the list of symbols seen throughout this paper.

\iftrue
\begin{table}[h!]
	\centering
	\caption{{List of Abbreviations (in alphabetic order)}}
	\begin{tabularx}{0.475\textwidth}{|lX|}\hline
		Abbreviation & Definition\\\hline
		AADRNN & Auto-Associative Deep Random Neural Network\\
		AAM & Auto-Associative Memory \\
		AE & Auto Encoder \\
		ANOVA & Analysis of Variance \\
		CDIS & Compromised Device Identification System \\
		CNN & Convolutional Neural Network\\
		DARPA & Defense Advanced Research Projects Agency\\	
		DDoS & Distributed Denial of Service \\
		DL & Deep Learning\\
		DNS & Domain Name System \\
		DoS & Denial of Service \\
		DRNN & Deep Random Neural Network \\ 
		DT & Decision Tree \\
		ELM & Extreme Learning Machine \\
		FISTA & Fast Iterative Shrinkage-Thresholding Algorithm\\
		IDS & Intrusion Detection System\\
		IoT	& Internet of Things\\
		IP & Internet Protocol\\
		ISSL & Incremental Semi-Supervised Learning\\
		KL & Kullback-Leibler \\
		KNN & K-Nearest Neighbour \\
		Lasso & Least Absolute Shrinkage and Selector Operator \\
		LR & Linear Regression \\
		LSTM & Long-Short Term Memory\\
		M2M & Machine-to-Machine\\
		MitM & Man-in-the-Middle \\
		ML & Machine Learning\\
		MLP & Multi-Layer Perceptron \\
		NB & Naive Bayes\\
		RF & Random Forest\\
		RNN & Random Neural Network \\
		ROC & Receiver Operating Characteristic \\
		SSID & Self-Supervised Intrusion Detection\\\hline
	\end{tabularx}
	\label{table:abbreviation_SSID}
\end{table}
\fi

\begin{table}[h!]
	\centering
	\caption{{List of Symbols}}	
	\begin{tabularx}{0.475\textwidth}{|lX|}
		\hline
		Symbol & Definition\\ \hline
		$M$ & Total number of network traffic metrics which is equivalent to the number of inputs to the IDS\\
		$x_{i}^{m}$ & The $m-th$ metric extracted from network traffic for packet $i$\\
		$x_i$ & The vector of input metrics, $x_i=[x_{i}^{1},~\cdots~ x_{i}^{m}, ~\cdots~,x_{i}^{M}]$\\
		$y_i$ & Binary output of the IDS indicating whether the packet $i$ is malicious\\
		$\hat{x}_i$ & The output of DL-based AAM indicating the expected value of $x_i$ in the absence of intrusion\\
		$W_h$ & Matrix of connection weights (including biases) between layer $h-1$ and $h$\\
		$W$ & Total number of learnable parameters in the ML model utilized in the IDS \\
		$\zeta(\cdot)$ & Activation function of a cluster in AADRNN\\
		$\Gamma$ & Trust coefficient indicating the trust of SSID on the decisions of the IDS\\
		$\Theta$ & Threshold on the trust coefficient (i.e. minimum desired value of $\Gamma$)\\
		$B^l$ & Batch of packets selected for learning\\
		$K$ & Minimum number of packets to be learned in each learning phase\\
		$I$ & Length of window in terms of the number of packets to calculate average decision\\
		$\gamma$ & Intrusion threshold \\
		$p^-_i$ & Probability of selecting packet $i$ to use as a benign packet\\
		$p^+_i$ & Probability of selecting packet $i$ to use as a malicious packet\\
		$q_i$ & Probability of rejecting packet $i$ to use in training\\
		$C_{rep}$ & Factor of representativeness of the traffic packets learned by IDS\\ 
		$C_{gen}$ & Factor of generalization ability of IDS\\ 
		$S^{TT}_{l}$ & Set of inter-transmission times of all packets learned by IDS until the end of learning phase $l$, which has mean of $1/\lambda_l$\\
		$S^{PL}_{l}$ & Set of packet lengths times of all packets learned by IDS until the end of learning phase $l$, which has mean of $1/\lambda_o$\\
		$S^{TT}_{o}$ & Set of inter-transmission times of all normal packets observed during continuous detection, which has mean of $1/\mu_l$\\
		$S^{PL}_{o}$ & Set of packet lengths times of all normal packets observed during continuous detection, which has mean of $1/\mu_o$\\
		$D^{TT}_{KL}$ & KL-Divergence from $S^{TT}_l$ to $S^{TT}_o$\\\vspace{-0.25cm}
		& \\
		$D^{PL}_{KL}$ & KL-Divergence from $S^{PL}_l$ to $S^{PL}_o$\\\vspace{-0.25cm}
		& \\
		$D^{TT}_{KL-norm}$ & Normalized KL-Divergence from $S^{TT}_l$ to $S^{TT}_o$\\\vspace{-0.25cm}
		& \\
		$D^{PL}_{KL-norm}$ & Normalized KL-Divergence from $S^{PL}_l$ to $S^{PL}_o$\\
		$\Delta$ & Adequacy of the packets learned by IDS \\
		$\kappa$ & Knowledge of IDS obtained from the packets learned\\
		$\mathcal{E}(l)$ & Empirical error measured at the end of learning phase $l$\\
		\hline
	\end{tabularx}
	\label{table:symbols_SSID}
\end{table}

\bibliographystyle{IEEEtran}
\bibliography{self_supervised,techniques,datasets,bot_detection,attack_detection,statistics,background,iot,security_issues_iot,RNN}

\begin{IEEEbiography}[{\includegraphics[width=1in,height=1.2in,clip,keepaspectratio]{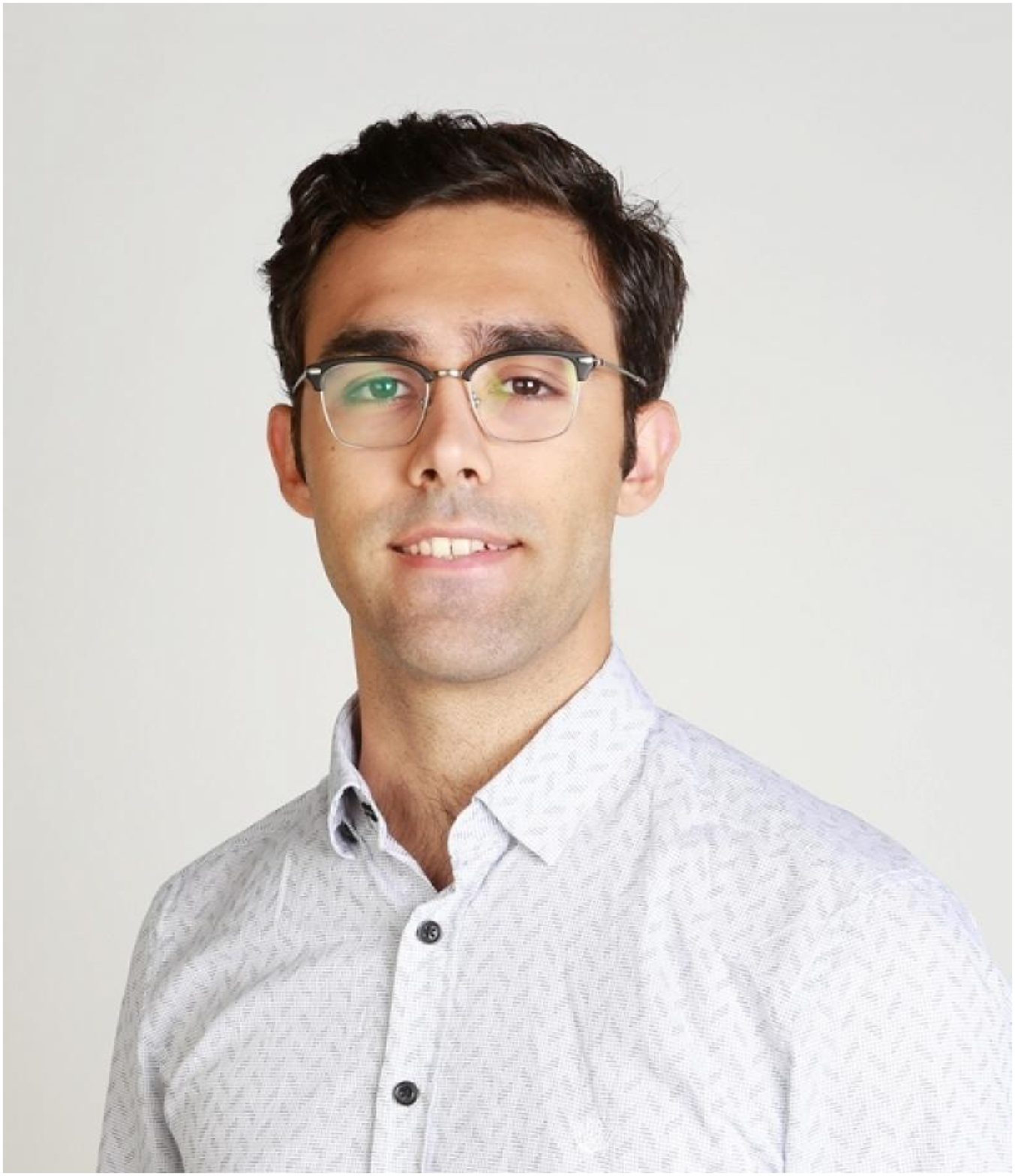}}]{\bf{Mert NAKIP}} obtained his B.Sc. degree, graduating with the first rank in his class, and subsequently completed his M.Sc. thesis in Electrical-Electronics Engineering at Yaşar University (Izmir, Turkey) in 2018 and 2020, respectively. His design of a multi-sensor fire detector utilizing ML achieved the national \#1 ranking at the Industry-Focused Undergraduate Graduation Projects Competition organized by TÜBİTAK (Turkish Scientific and Technological Research Council). His M.Sc. thesis, focusing on applying machine learning techniques to Massive Access of IoT, supported by the National Graduate Scholarship Program of TÜBİTAK 2210C in High-Priority Technological Areas. He received his Ph.D. from the Institute of Theoretical and Applied Informatics, Polish Academy of Sciences (Gliwice, Poland) in January 2024, and is now serving as an Assistant Professor there. He has participated in projects funded by TÜBİTAK and the European Commission concerning IoT, cybersecurity, and machine learning. Currently, he is involved in the DOSS project.	
\end{IEEEbiography}

\begin{IEEEbiography}
	[{\includegraphics[width=1in,height=1.2in,clip,keepaspectratio]{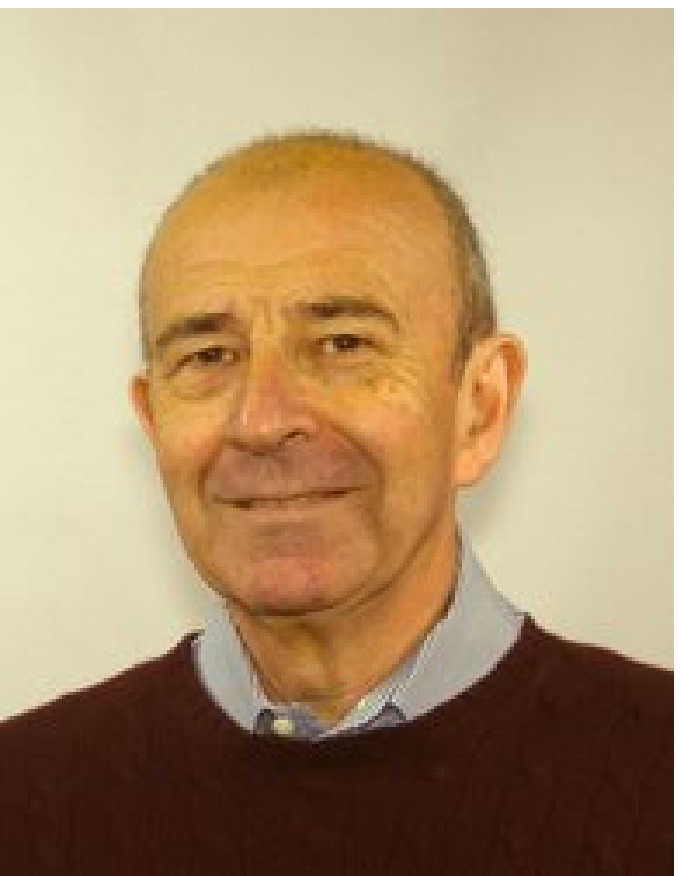}}]{\bf Erol GELENBE} FIEEE, FACM, FIFIP, FRSS, FIET, graduated from METU (Turkey), and received the PhD from Polytechnic Institute of NYU, and the DSc from Sorbonne University. He pioneered system performance evaluation methods, and invented the Random Neural Network, while helping to develop commercial products including the Queueing Network Analysis Package and the manufacturing simulator FLEXSIM. He has graduated over 90 PhDs including 94 women. Professor at the Institute of Theoretical and Applied Informatics, Polish Academy of Sciences, he previously held chairs at Imperial College, University of Central Florida, Duke University, Université Paris-Descartes and Paris-Saclay, and Université de Liège. His prizes include the Parlar Foundation Science Award (Turkiye), Grand Prix France Télécom, the ACM SIGMETRICS Life-Time Achievement Award, the UK IET Oliver Lodge Medal and the Mustafa Prize. Elected Fellow of Academia Europaea, the French National Acad. of Technologies, the Science Academies of Hungary, Poland, Turkey and the Royal Acad. of Belgium. He was awarded Chevalier de la Légion d’Honneur, Chevalier des Palmes Académiques and Commandeur du Mérite by France, Commander of the Order of the Crown of Belgium, Commendatore al Merito and Grande Ufficiale dell’Ordine della Stella by Italy, and Officer of the Order of Merit of Poland. Principal Investigator of numerous European Union research projects, Coordinator of FP7 NEMESYS and H2020 SerIoT, he has also been supported by the US NSF, ONR, ARO, the UKRI and industry.
\end{IEEEbiography}

\end{document}